\documentclass{revtex4}
\usepackage{graphicx,mathrsfs}
\begin{document}
\title{Waves statistics for generalized one-dimensional Nonlinear Schrodinger Equation with saturated nonlinearity}

\author{D.S. Agafontsev$^{(a),(b)}$}
\affiliation{\small \textit{ $^{(a)}$ P. P. Shirshov Institute of Oceanology, 36 Nakhimovsky prosp., Moscow 117218, Russia.\\
$^{(b)}$ Novosibirsk State University, 2 Pirogova, 630090 Novosibirsk, Russia.}}

\begin{abstract}
We measure spectra, spatial correlation functions and probability density functions (PDFs) for waves amplitudes for generalized one-dimensional nonlinear Schrodinger (NLS) equation of focusing type with saturated nonlinearity. All additional terms beyond the classical NLS equation are small. As initial data we use perturbed by weak noise modulationally unstable condensate. On the PDFs we observe power-law region $PDF(|\Psi|)\sim |\Psi|^{-1}$ for small $|\Psi|\ll\sqrt{\langle|\Psi|^{2}\rangle}$ and medium $|\Psi|\sim\sqrt{\langle|\Psi|^{2}\rangle}$ amplitudes followed by intermediate region and then Rayleigh far tail. Power-law region appears starting from some critical levels of average amplitude $\sqrt{\langle|\Psi|^{2}\rangle}$ and coefficient $\alpha$ related to saturated nonlinearity, and then becomes more pronounced with $\sqrt{\langle|\Psi|^{2}\rangle}$ and $\alpha$. Correlation of phases becomes significant for large wave events and contributes about one order of magnitude to the frequencies of their 
occurrence. Waves statistics for the 
considered system turns out to be exceptionally stable against additional stochastic forces.
\end{abstract}

\maketitle


{\bf 1.} Statistics of waves for different nonlinear systems has drawn much scientific attention in the recent time \cite{Dudley1, Dudley2, Hadzievski, Taki1, Bortolozzo, Dudley3, Taki2, Lushnikov, Agafontsev2}, especially since the first observation of optical rogue waves \cite{Solli} - large wave events that appear randomly from initially smooth pulses and have statistics drastically different from that predicted by the linear theory. One of the scenarios for rogue waves appearance in optics and also hydrodynamics is realized via nonlinear focusing of waves during the modulation instability development (see \cite{Solli, Kharif, Dysthe}) that is described by the classical nonlinear Schrodinger (NLS) equation of focusing type, 
$$
i\Psi_t +\beta\Psi_{xx}+\gamma|\Psi|^2 \Psi = 0,
$$
starting from the initial condensate state 
$$
\Psi(t=0) = C+\epsilon(x)
$$ 
where $C$ is constant, $|\epsilon(x)|\ll|C|$ is a small noise, $t$ is time, $x$ is spacial coordinate, $\beta$ and $\gamma$ are real nonzeroth coefficients such that $\beta\gamma>0$, and $\Psi$ is wave field or wave field envelope. After the scaling and gauge transformations $x=\tilde{x}\sqrt{\beta/(\gamma|C|^{2})}$, $t=\tilde{t}/(\gamma|C|^{2})$, $\Psi=C\tilde{\Psi}e^{i\tilde{t}}$ and $\epsilon=C\tilde{\epsilon}e^{i\tilde{t}}$, this problem is reduced to
\begin{equation}\label{Eq01}
i\Psi_t -\Psi +\Psi_{xx}+|\Psi|^2 \Psi = 0,\quad\quad \Psi(t=0) = 1+\epsilon(x),
\end{equation}
where all tilde signs are omitted. In the framework of Eq. (\ref{Eq01}) modulation instability develops on the background of the exact condensate solution $\Psi=1$, amplifying small periodic modulations 
$$
\Psi = 1 + \kappa\exp(ikx+i\Omega t),\quad\quad \Omega^{2}=k^{4}-2k^{2},
$$
for wavenumbers $k\in (-\sqrt{2}, \sqrt{2})$, and the maximum increment of the instability is realized at $|k|=k_{0}=1$.

Nonlinear term $|\Psi|^2 \Psi$ in Eq. (\ref{Eq01}) appears in optics because of Kerr nonlinearity - dependence of refraction index on wave amplitude $E$ in the form $n=n_{0}+n_{2}|E|^{2}$. Refraction index cannot grow to infinity with intensity $|E|^{2}$, for large intensities of order 1 $GW/cm^{2}$ saturation effects come into play (see \cite{Agrawal, Mihalache} and also \cite{Max} for plasma waves) and Eq. (\ref{Eq01}) is modified as
\begin{equation}\label{Eq02}
i\Psi_t - \Psi + \Psi_{xx}+\frac{|\Psi|^2}{1+\alpha|\Psi|^2}\Psi = 0,
\end{equation}
or
\begin{equation}\label{Eq03}
i\Psi_t - \Psi + \Psi_{xx}+\frac{1-\exp(-\alpha|\Psi|^2)}{\alpha}\Psi = 0, 
\end{equation}
where $\alpha>0$ is saturation parameter and equations (\ref{Eq02}) and (\ref{Eq03}) correspond to different dielectric tensors (see \cite{Mihalache}).

\begin{figure}[h]\centering
\includegraphics[width=130pt]{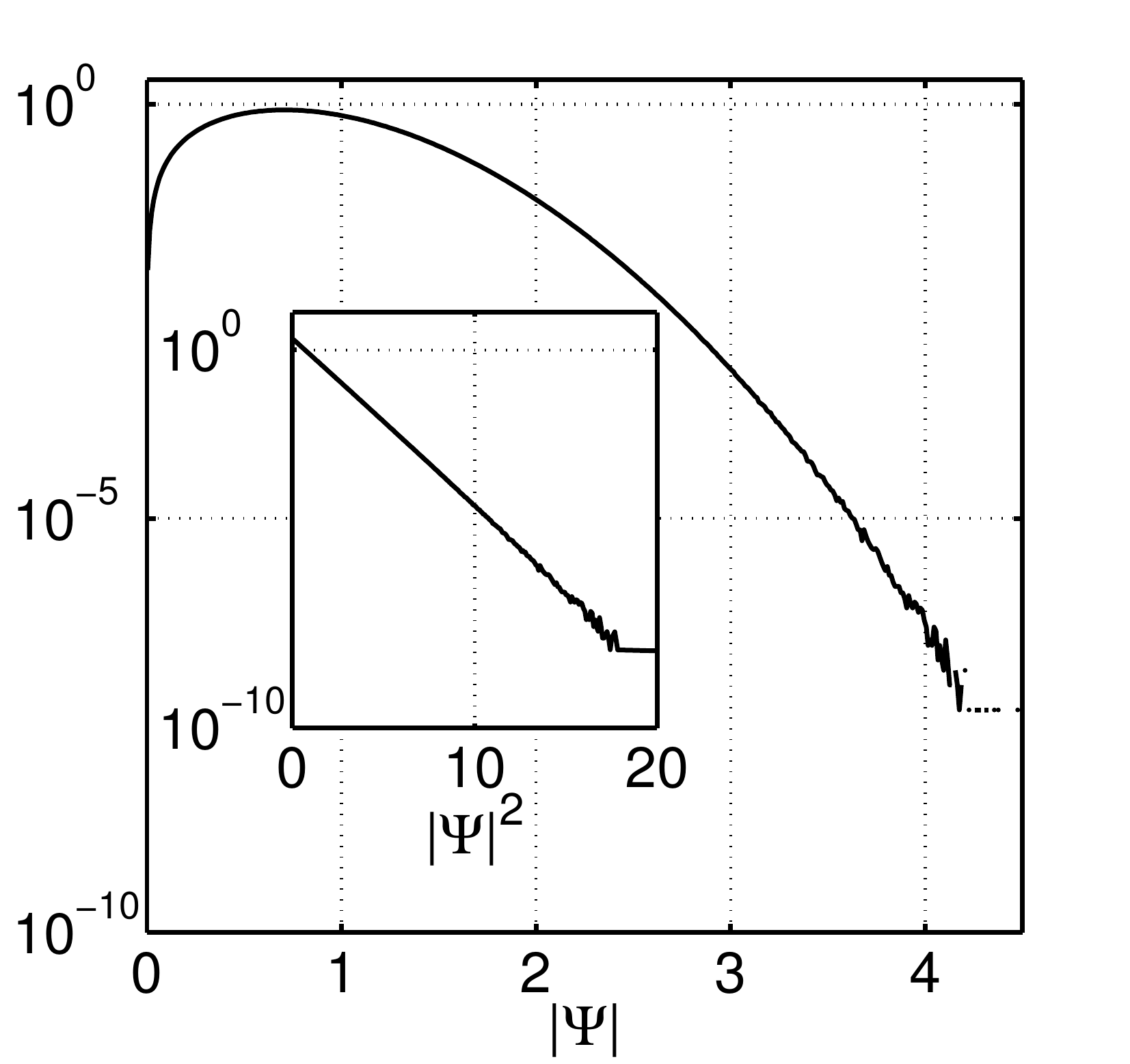}

\caption{\small {\it $PDF(|\Psi|)$ depending on $|\Psi|$ for linear waves $\Psi(x) = A\int_{-\infty}^{+\infty}\exp(-k^{2}/\theta^{2} + i\phi_{k})e^{ikx}(dk/2\pi)$, $\theta=5$, $A=10$, calculated using $10^{6}$ different realizations of random uncorrelated phases $\phi_{k}$. Inset shows $PDF(|\Psi|)/|\Psi|$ depending on $|\Psi|^{2}$.}}
\label{fig:linearPDF}
\end{figure}

Let us suppose that the current state of a system consists of multitude of uncorrelated linear waves,
\begin{equation}\label{DFT}
\Psi = \sum_{k}a_{k}\, \exp(i(kx-\omega_{k}t+\phi_{k})).
\end{equation}
If $a_{k}$ and $\phi_{k}$ are random uncorrelated values and the number of waves $\{k\}$ is large enough, then under the conditions of central limit theorem real $Re\,\Psi$ and imaginary $Im\,\Psi$ parts of field $\Psi$ are Gaussian-distributed and probability to meet amplitude $|\Psi|$ (probability density function, PDF) obeys Rayleigh distribution (see example on FIG.~\ref{fig:linearPDF}),
\begin{equation}\label{Rayleigh}
PDF(|\Psi|) \sim |\Psi|\exp(-|\Psi|^{2}/2\sigma^{2}).
\end{equation}

In the recent publication \cite{Agafontsev2} PDFs for waves amplitudes were analyzed for modulation instability development described by the classical NLS equation of focusing type (\ref{Eq01}). For this analysis large ensembles of initial distributions $\Psi(t=0)=1+\epsilon(x)$, where $|\epsilon(x)|\ll 1$ is small initial space-homogeneous noise, were taken and with the help of numerical simulations their evolution with time $\Psi(t,x)$ was studied. Initial distributions $\Psi(t=0)$ differed only by realizations of noise $\epsilon(x)$ with fixed noise statistical properties inside each of the ensembles. Based on these simulations it was shown that PDFs for waves amplitudes for problem (\ref{Eq01}) generally are still very similar to Rayleigh ones, with small time-dependent deviations in the region of medium amplitudes.

In the current publication we study the same scenario of modulation instability development but in the framework of generalized one-dimensional NLS equation with saturated nonlinearity (\ref{Eq02}) or (\ref{Eq03}). For this purpose we measure energy spectrum $I_{k}=\langle|\Psi_{k}|^{2}\rangle$ (here and below $\langle..\rangle$ stands for averaging over ensemble and $\Psi_{k}$ is Fourier transform of $\Psi(x)$), spatial correlation functions $g(x)=\langle\Psi(y,t)\Psi^{*}(y+x,t)\rangle$ and the PDFs for waves amplitudes. We use term "PDF" only in relation to PDFs for waves amplitudes. Since the classical NLS equation with small additional terms is a very common model in physics, we consider only those systems with saturated nonlinearity that have dynamics close to that of the classical NLS equation, limiting ourselves with small saturation parameters $\alpha \ll 1$.

Because of the equality $\int F(x)x\,dx = (1/2)\int F(x)\,d\,x^{2},$ a PDF for squared amplitudes $|\Psi|^{2}$, that by definition is the probability to meet a given squared amplitude $|\Psi|^{2}$, is exponential if the corresponding amplitude PDF is Rayleigh one,
$$
PDF(|\Psi|) \sim |\Psi|\exp(-|\Psi|^{2}/2\sigma^{2})\quad \Leftrightarrow\quad PDF(|\Psi|^{2}) \sim \exp(-|\Psi|^{2}/2\sigma^{2}),
$$
and vice versa. It is more convenient to examine exponential $\exp(-z), \, z=|\Psi|^{2},$ dependencies than Rayleigh $z\exp(-z^{2}), \, z=|\Psi|,$ ones, therefore, as in \cite{Agafontsev2}, in this publication we measure PDFs for squared amplitudes $|\Psi|^{2}$ instead of PDFs for amplitudes $|\Psi|$ and compare the results with exponential dependencies that we call Rayleigh ones for simplicity. If not stated otherwise, we measure squared amplitude PDFs for entire field $\Psi$ (in contrast to local maximums or absolute maximums PDFs) and use normalization,
$$
\int_{0}^{+\infty} PDF(|\Psi|^{2})\,d|\Psi|^{2}=1.
$$

The paper is organized as follows. The next section gives overview of nonlinear systems we examine and also numerical methods we use. Results of our numerical simulations are presented in Section 3. Section 4 contains conclusions and acknowledgements.\\


{\bf 2.} For sufficiently small saturation parameters and amplitudes $a|\Psi|^{2}\ll 1$ Eq. (\ref{Eq02}) and (\ref{Eq03}) can be expanded in series with respect to powers of amplitude $|\Psi|$,
$$
i\Psi_t - \Psi + \Psi_{xx}+\Psi(|\Psi|^2 - \alpha|\Psi|^4 + \alpha^{2}|\Psi|^6 - ...) = 0,
$$
and
$$
i\Psi_t - \Psi + \Psi_{xx}+\Psi(|\Psi|^2 - (\alpha/2)|\Psi|^4 + (\alpha^{2}/6)|\Psi|^6 - ...) = 0,
$$
respectively. Our test simulations demonstrated that for comparatively small saturation parameters $\alpha \sim 0.1$ saturated nonlinearity of Eq. (\ref{Eq03}) provides results for the statistics of waves almost indistinguishable from that for saturated nonlinearity of Eq. (\ref{Eq02}) with twice less saturation parameters $\alpha/2$. This means that for the considered systems waves statistics is mainly defined by the defocusing six-wave interactions while the influence of eight-wave and higher-order interactions is small. Below we will discuss saturated nonlinearity of Eq. (\ref{Eq02}) only.

Eq. (\ref{Eq02}) is the Hamiltonian one,
$$
i\Psi_t = \frac{\delta H}{\delta \Psi^{*}},
$$
with Hamiltonian
\begin{equation}
H = E + N,\quad\quad E = H_{d} + H_{n},
\end{equation}
where $N=\int|\Psi|^{2}\,dx$ is wave action, $E$ is total energy, $H_{d}$ is kinetic energy,
\begin{equation}
H_{d} = \int |\Psi_{x}|^2\,dx,
\end{equation}
and $H_{n}$ is potential energy
\begin{equation}
H_{n} = \frac{1}{\alpha^{2}}\int \bigg(\ln(1+\alpha|\Psi|^2) - \alpha|\Psi|^2\bigg)\,dx = \int\bigg(-\frac{|\Psi|^4}{2}+\frac{\alpha|\Psi|^6}{3}-\frac{\alpha^{2}|\Psi|^8}{4}+...\bigg)\,dx.
\end{equation}
It will be informative to further subdivide potential energy $H_{n}$ by the energy of four-wave interactions,
\begin{equation}
H_{4} = -\int\frac{|\Psi|^4}{2}\,dx,
\end{equation}
and the energy of higher-order nonlinear interactions,
\begin{equation}
H_{6} = H_{n}-H_{4}.
\end{equation}
In addition to Hamiltonian $H$ and total energy $E$, Eq. (\ref{Eq02}) also conserves wave action $N$ and momentum $P=(i/2)\int(\Psi_{x}^{*}\Psi-\Psi_{x}\Psi^{*})\,dx$. The classical NLS equation can be obtained from Eq. (\ref{Eq02}) in the limit $\alpha\to 0$, and due to complete integrability in terms of inverse scattering transformation it conserves an infinite number of integrals of motion where wave action, momentum and Hamiltonian $H=H_{d}+H_{4}$ are the first three ones.

However, direct investigation of waves statistics in the framework of Eq. (\ref{Eq02}) turns out to be not very informative since the system exhibits relaxation phenomena - statistically irreversible movement to it's statistical attractor in the form of one big soliton containing all the potential energy and immersed in the small fluctuations field (same for Eq. (\ref{Eq03}) - see \cite{Jordan} for both systems). And even though for small time shifts the dynamics of Eq. (\ref{Eq02}) resembles that of the classical NLS equation, for larger time shifts it demonstrates significantly different behavior. In particular, our test simulations revealed that after some time energy of higher-order nonlinear interactions $H_{6}$ becomes comparable with energy of four-waves interactions $H_{4}$ and kinetic energy $H_{d}$. At this point the dynamics of Eq. (\ref{Eq02}) can no longer be approximated by the classical NLS equation. 

It is also necessary to mention another important circumstance. At very large time shifts the PDFs for Eq. (\ref{Eq02}) represent amplitudes distribution of the final big solitons, while the parameters of these final solitons can be directly calculated from the values of three integrals of motion - wave action, momentum and Hamiltonian \cite{Jordan}. These values depend on the length of integration region, therefore the PDFs for pure Eq. (\ref{Eq02}) become dependent on numerical integration parameters.

In order to overcome such limitations we introduce dumping in the form of linear dissipation ($-id_{l}\Psi_{xx}$), two- ($id_{2p}|\Psi|^2\Psi$) and three-photon absorption ($id_{3p}|\Psi|^4\Psi$) terms, and also a general pumping term $\Phi$, the similar way as it was done in \cite{Agafontsev2, Lushnikov}:
\begin{equation}\label{Eq021}
i\Psi_t - \Psi + (1-id_{l})\Psi_{xx}+\frac{|\Psi|^2}{1+\alpha|\Psi|^2}\Psi + id_{2p}|\Psi|^2\Psi + id_{3p}|\Psi|^4\Psi = i\Phi,
\end{equation}
where $d_{l}$, $d_{2p}$ and $d_{3p}$ are small positive constants:
$$
d_{l},d_{2p},d_{3p} > 0, \quad d_{l},d_{2p},d_{3p} \ll 1. 
$$

Although one of the reasons for the addition of dumping and pumping terms is to hold Eq. (\ref{Eq02}) from relaxation, these terms also have clear physical meaning. Saturation of nonlinearity becomes significant at very high amplitudes that in optics makes important also two- and three-photon absorption terms. Through combination of nonlinearity and dispersion big amplitudes may lead to pronounced widening of spectra for which it is necessary to take into account linear filtering term $-id_{l}\Psi_{xx}$. The specific form of the pumping term $i\Phi$, however, depends on the physical model. For some systems (waves in plasmas and fluids, Josephson junctions, some optical problems including lasers far from saturation energies - see \cite{Lushnikov2, Lushnikov21, Lushnikov22, Lushnikov23, Turitsyn}) the pumping term may be represented as 
\begin{equation}\label{Eq041}
\Phi_{1}=\hat{p}\Psi,
\end{equation}
where $\hat{p}$ is a linear integral operator, so that in k-space $\Phi_{k}=p_{k}\Psi_{k}$. Here we limit ourselves with consideration of k-independent pumping term $p_{k}=p_{1}>0$ only. For other systems additive random forcing might be important, here we model it as space-homogeneous superposition of Gaussian-distributed in k-space linear waves
\begin{equation}\label{Eq051}
\Phi_{2}(x,t)=p_{2} \int \exp(-k^{2}/\theta_{p}^{2}+i\xi_{k}(t))\,e^{ikx}\,\frac{dk}{2\pi},
\end{equation}
with some coefficient $p_{2}>0$, relatively large dispersion $\theta_{p}\gg k_{0}$ ($k_{0}=1$ corresponds to the maximum growth rate of the modulation instability; $\theta_{p}=5$ for most of the simulations) and arbitrary phases $\xi_{k}(t)$ for each $t$, so that $\Phi_{2}(x,t)$ is $\delta$-correlated in time and Gaussian-correlated in space:
$$
\langle \Phi_{2}(x_{1},t_{1})\Phi_{2}^{*}(x_{2},t_{2})\rangle_{\xi} = D\delta(t_{2}-t_{1})\exp(-(x_{2}-x_{1})^{2}/\Delta^{2}),
$$
where $\langle..\rangle_{\xi}$ stands for averaging over realizations of phases $\xi_{k}$. Further we will refer to average squared amplitude of the stochastic pumping term $\Phi_{2}$ that can be calculated as follows,
\begin{eqnarray}\label{Eq061}
&&\langle|\Phi_{2}|^{2}\rangle_{\xi}=\frac{\langle\int |\Phi_{2}|^{2}\, dx\rangle_{\xi}}{\int dx}=\nonumber\\
&&=\frac{p_{2}^{2}}{L}\bigg\langle\int \exp[-(k_{1}^{2}+k_{2}^{2})/\theta_{p}^{2}+i(\xi_{k_{1}}-\xi_{k_{2}})]e^{i(k_{1}-k_{2})x}\, \frac{dk_{1}dk_{2}}{(2\pi)^{2}}dx\bigg\rangle_{\xi} = \frac{1}{\sqrt{8\pi}}\frac{\theta_{p}}{L}p_{2}^{2},
\end{eqnarray}
where $L=\int dx$ is length of the region of integration. In the current publication we use deterministic pumping term $\Phi=p_{1}\Psi$, and then study the influence of stochastic pumping superimposed over deterministic one $\Phi = p_{1}\Psi + \Phi_{2}$.

During the evolution of wave field $\Psi$ in the framework of Eq. (\ref{Eq021}) one-dimensional wave turbulence is developed. In the integrable case (\ref{Eq01}) the turbulence is called integrable and relaxes to one of infinite possible stationary states. In case of Eq. (\ref{Eq021}) wave field $\Psi$ comparatively quickly (tens of nonlinear lengths) approaches to the statistically steady state when energy drain due to dissipation is compensated by energy income due to pumping and wave action, momentum and total energy as well as kinetic $H_{d}$, four-wave interactions $H_{4}$ and higher-order interactions energy $H_{6}$ fluctuate near their mean values. Tuning of dumping and pumping parameters allows one to reach such statistically steady states that $|H_{d}|\sim |H_{4}|$, and $|H_{6}|\ll |H_{d}|, |H_{4}|$, i.e. when the dynamics of Eq. (\ref{Eq021}) resembles that of the classical NLS equation. These statistically steady states can also be described as quasi-solitonic turbulence (see \cite{Zakharov1, 
Zakharov2}) when quasi-solitons play significant role in the dynamics and re-distribution of energy inside the systems. 

Most frequently we will use the following saturation, dumping and pumping parameters:
\begin{equation}\label{parameters}
\alpha=0.04,\quad d_{l}=0.0324,\quad d_{2p}=0,\quad d_{3p}=0.0002,\quad p_{1}=0.02,
\end{equation}
that are very similar to that used in \cite{Agafontsev2}. Dependence of the statistics of waves on these parameters will be examined below. Dumping and pumping coefficients in (\ref{parameters}) are sufficient to keep the system from relaxation, and in the corresponding statistically steady state the mean values for wave action, total energy, kinetic and four-wave interactions energy are close to that of the classical NLS equation.

We solve Eq. (\ref{Eq021}) numerically in the box $-16\pi \le x<16\pi$ with periodic boundary conditions, so that modulation instability generates 16 peaks in its initial stage. We start from initial data $\Psi|_{t=0}=1+\epsilon(x)$ where $|\epsilon(x)|\ll 1$ is stochastic noise,
$$
\epsilon(x)=A_{0} \int \exp(-k^{2}/\theta^{2}+i\xi_{k})\,e^{ikx}\,\frac{dk}{2\pi},
$$
with large dispersion $\theta>>1$ ($\theta=5$ for most of the simulations)  and arbitrary phases $\xi_{k}$ (compare with (\ref{Eq051})). We use coefficient $A_{0}=10^{-3}$ corresponding to average noise amplitude $\sqrt{\langle|\epsilon(x)|^{2}\rangle_{\xi}}\approx 10^{-4}$. We did not find significant difference in our results using other homogeneous in x-space statistical distributions of noise or other parameters $A_{0}$ and $\theta$.

In our numerical simulations we use 2nd-order Split-Step method when linear and nonlinear parts of the equations are calculated separately. In order to improve simulations and save computational resources we employ adaptive change of spacial grid size $\Delta x$ reducing it when Fourier components of solution $\Psi_{k}$ at high wave numbers $k$ exceed $10^{-13}\max|\Psi_{k}|$ and increasing $\Delta x$ when this criterion allows. In order to prevent appearance of numerical instabilities, time step $\Delta t$ also changes with $\Delta x$ as $\Delta t = h\Delta x^{2}$, $h \le 0.1$ (see \cite{Lakoba}). For most of the simulations we use ensembles of 10000 initial distributions each. We checked our statistical results obtained with the help of this numerical schema against the size of the ensembles and implementation of other numerical methods (like Runge-Kutta 4th and 5th order methods) and found no difference.\\


\begin{figure}[h] \centering
\includegraphics[width=130pt]{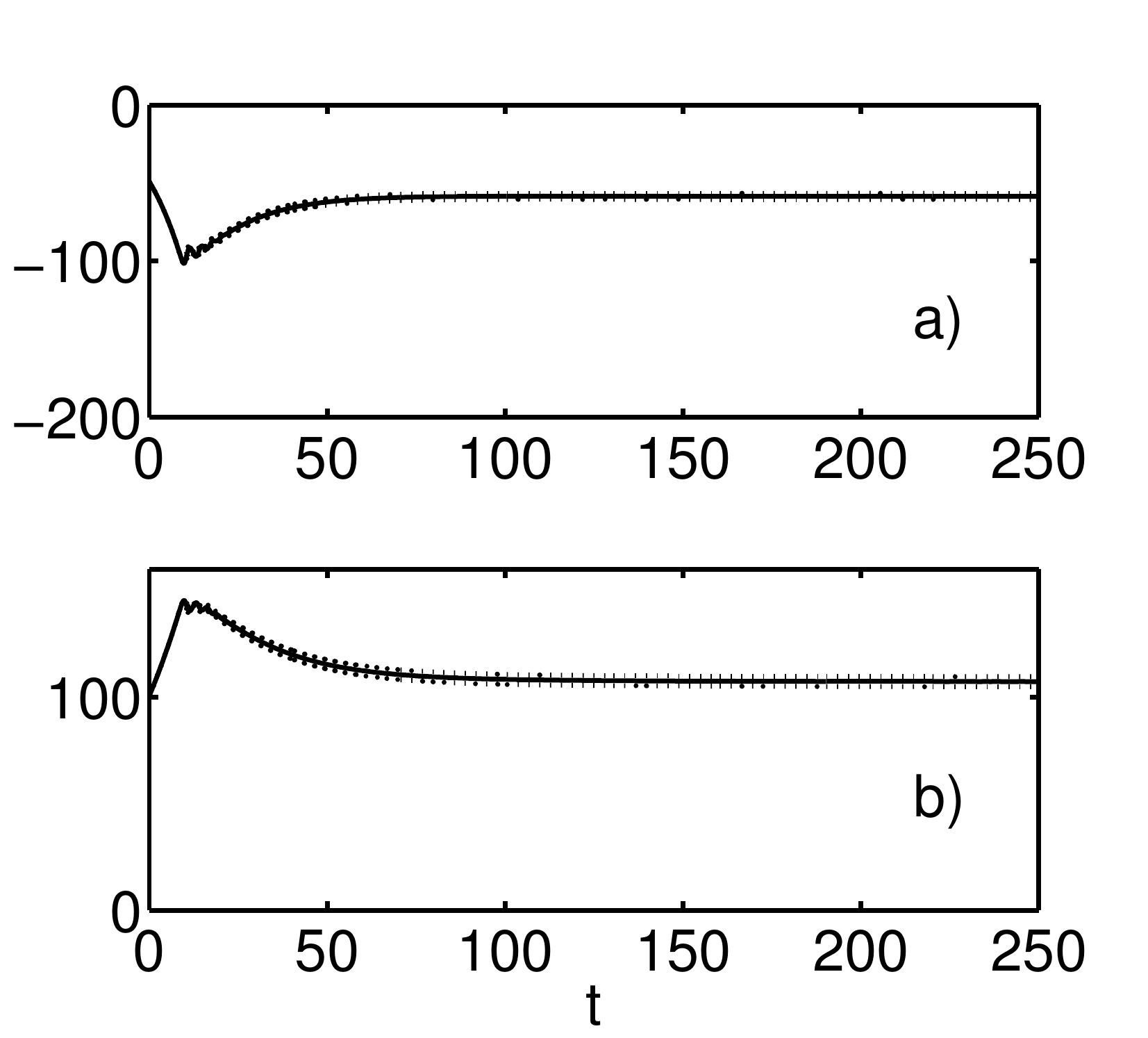}
\includegraphics[width=130pt]{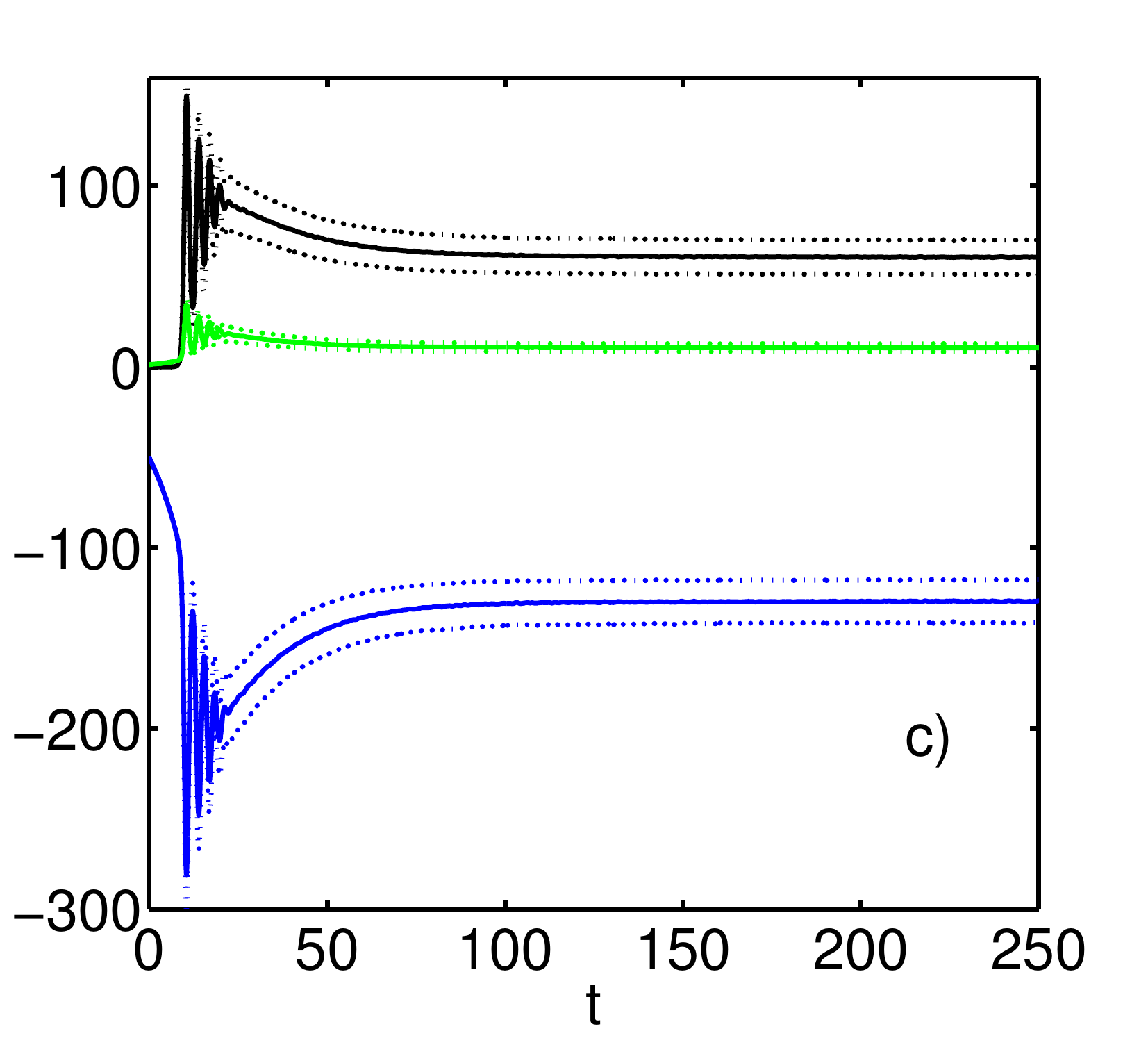}
\includegraphics[width=130pt]{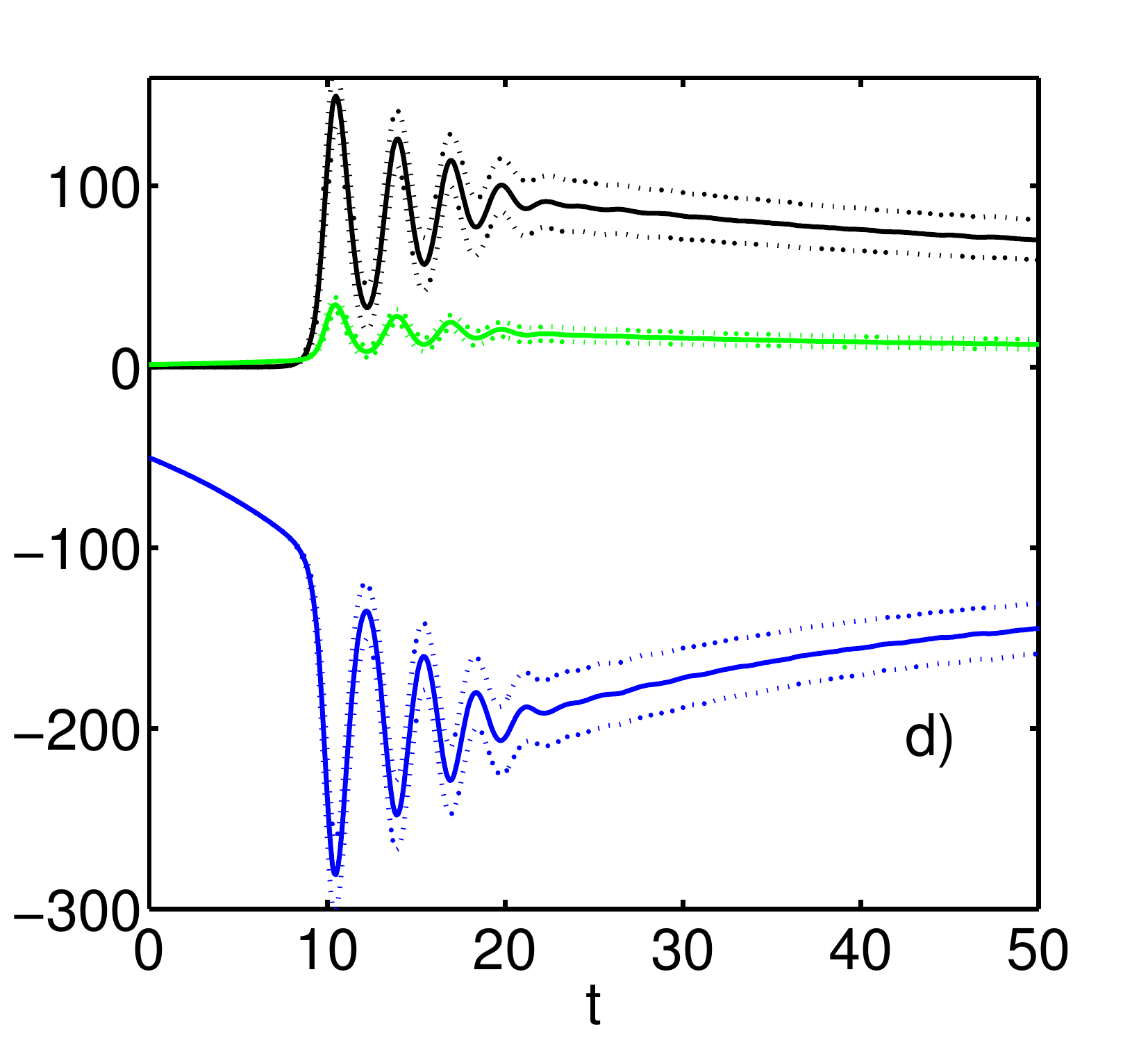}

\caption{\small {\it  (Color on-line) Evolution of averaged over ensemble (a) total energy $\langle E\rangle$, (b) wave action $\langle N\rangle$ and (c), (d) kinetic energy $\langle H_{d}\rangle$ (black), four-wave interactions $\langle H_{4}\rangle$ (blue) and higher-order interactions energy $\langle H_{6}\rangle$ (green) for Eq. (\ref{Eq021}) with parameters (\ref{parameters}). Solid lines - mean over ensemble values, dashed lines - borders for the corresponding standard deviations. Graph (d) is graph (c) enlarged at $t\in [0, 50]$.}}
\label{fig:EnergyEvolution}
\end{figure}

{\bf 3.} In the presence of dumping and pumping terms wave action $N$ and total energy $E$ for Eq. (\ref{Eq021}) become dependent on time; their evolution and also evolution of kinetic $H_{d}$, four-wave interactions $H_{4}$ and higher-order interactions $H_{6}$ energy is shown on FIG.~\ref{fig:EnergyEvolution}a,b,c - all averaged over ensemble. If not stated otherwise here and below we use deterministic pumping term $\Phi=p_{1}\Psi$.

While wave field $\Psi(x,t)$ is close to the condensate state $\Psi=1$, dispersion $\Psi_{xx}$ and linear dumping term $-id_{l}\Psi_{xx}$ are negligible because the spectrum $\Psi_{k}$ is concentrated in the zeroth harmonic $\Psi_{k=0}$ and all other harmonics are very small. As we start our simulations from the condensate state peturbed by weak random noise $\Psi(t=0)=1+\epsilon(x)$, these conditions are satisfied approximately up to time shifts $t<7$ for parameters (\ref{parameters}), and Eq. (\ref{Eq021}) can be approximated where by the linear equation
$$
i\Psi_t = \bigg(1-\frac{1}{1+\alpha}+i(p_{1}-d_{2p}-d_{3p})\bigg)\Psi,
$$
that for our choice of parameters $d_{2p}+d_{3p}\ll p_{1}$ means exponential uniform amplitude growth with time from $|\Psi|\approx 1$ at $t=0$ to $|\Psi|\approx 1.14$ at $t=7$. Corresponding to this process straight regions are clearly seen on FIG.~\ref{fig:EnergyEvolution}a,b,d. for total energy $E$, wave action $N$ and four-wave interactions energy $H_{4}$. Then linear dumping significantly increases and from $t>10$ pumps the excess of energy out of the system.

Net pumping or dumping of energy is accompanied by the modulation instability that becomes noticeable starting from $t>7$ and develops the same way as for the classical NLS equation. Thus, we observe on FIG.~\ref{fig:EnergyEvolution}d the same regular oscillations of kinetic $H_{d}$ and four-wave interactions $H_{4}$ energy as for the integrable case (\ref{Eq01}) (see \cite{Agafontsev2}) with the exception that modulation instability develops from slightly higher amplitudes. These oscillations are about three times higher than the corresponding standard deviations, and it is noteworthy that their influence on the evolution of wave action and total energy is very small. Oscillations cease to $t\sim 22$ and the system approaches at $t\sim 100$ to the statistically steady state when wave action, total energy, kinetic, four-wave interactions and higher-order interactions energy as well as spectra, spatial correlation functions, and the PDFs only slightly fluctuate with time (compare with \cite{Agafontsev2}). 
The latter allows us to perform additional averaging over time for these statistical characteristics.

\begin{figure}[h] \centering
\includegraphics[width=130pt]{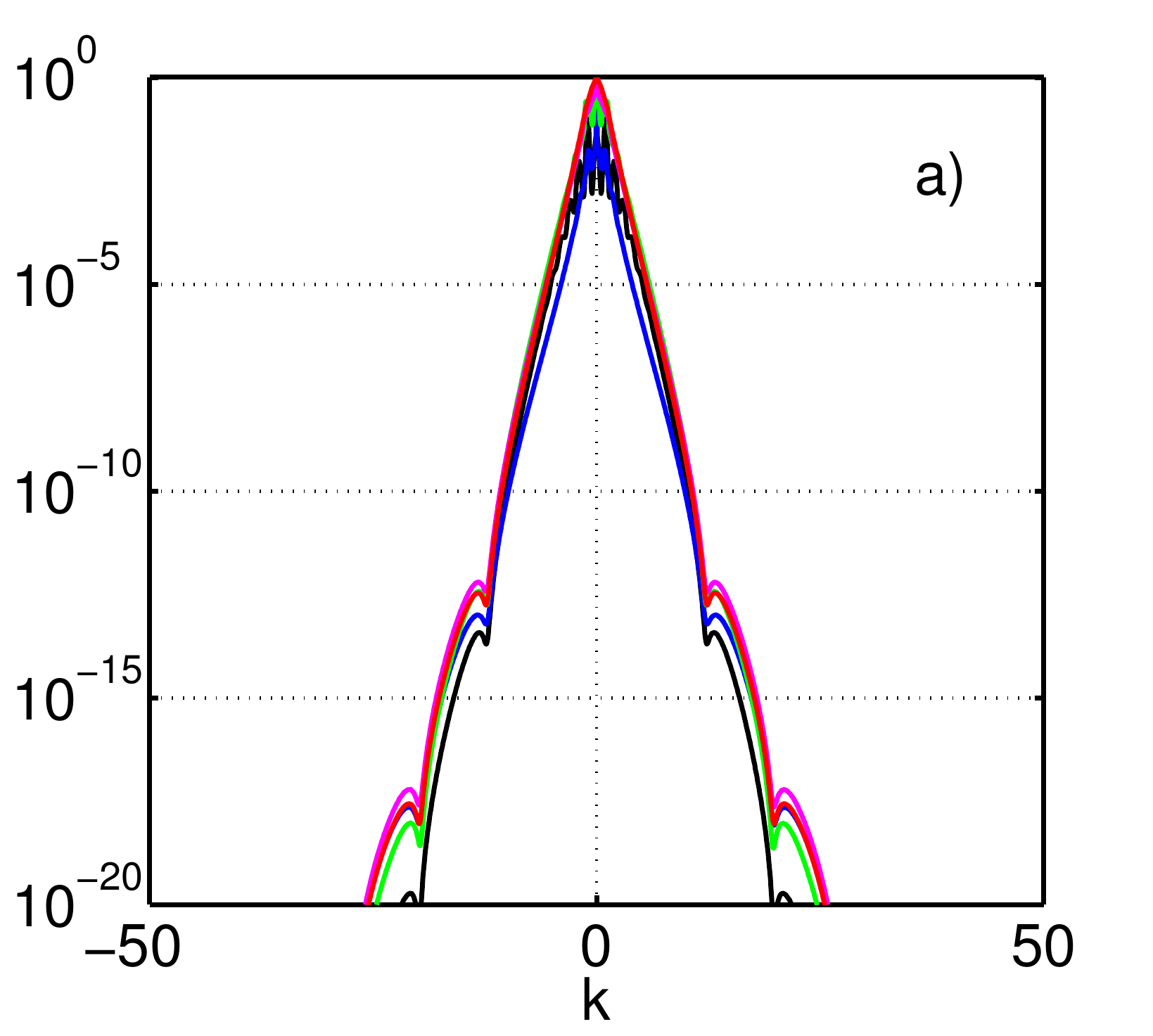}
\includegraphics[width=130pt]{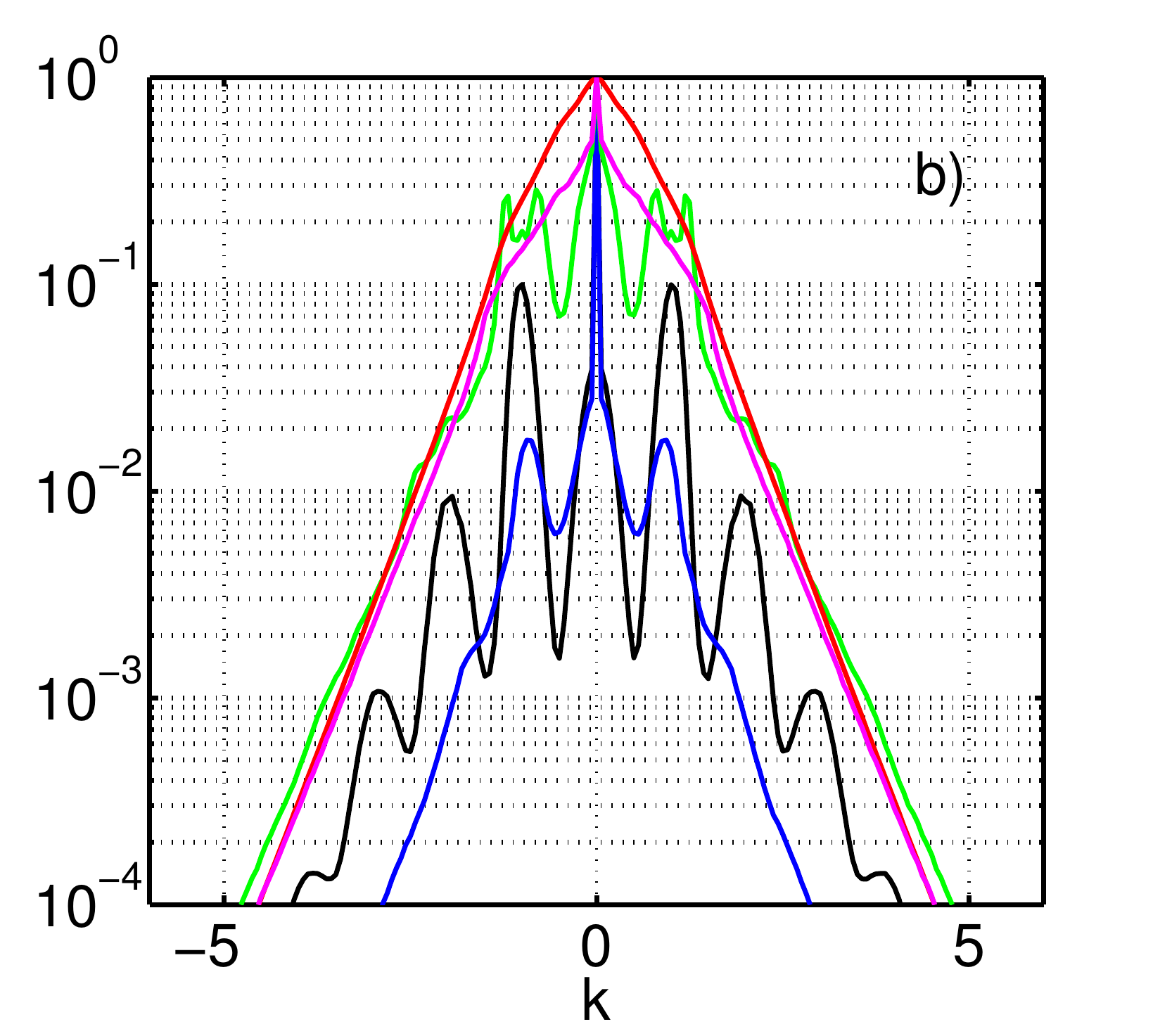}\\
\includegraphics[width=130pt]{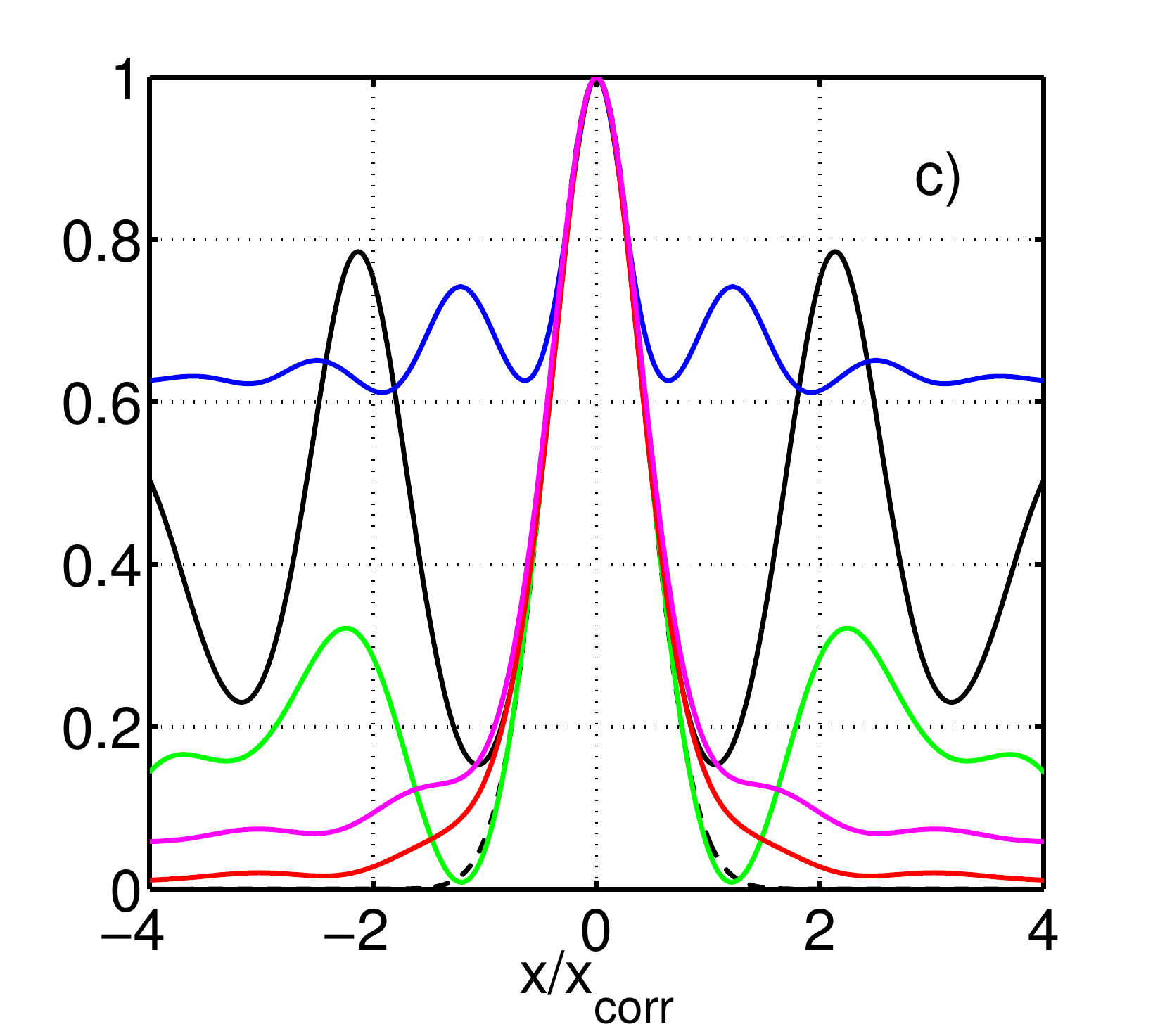}
\includegraphics[width=130pt]{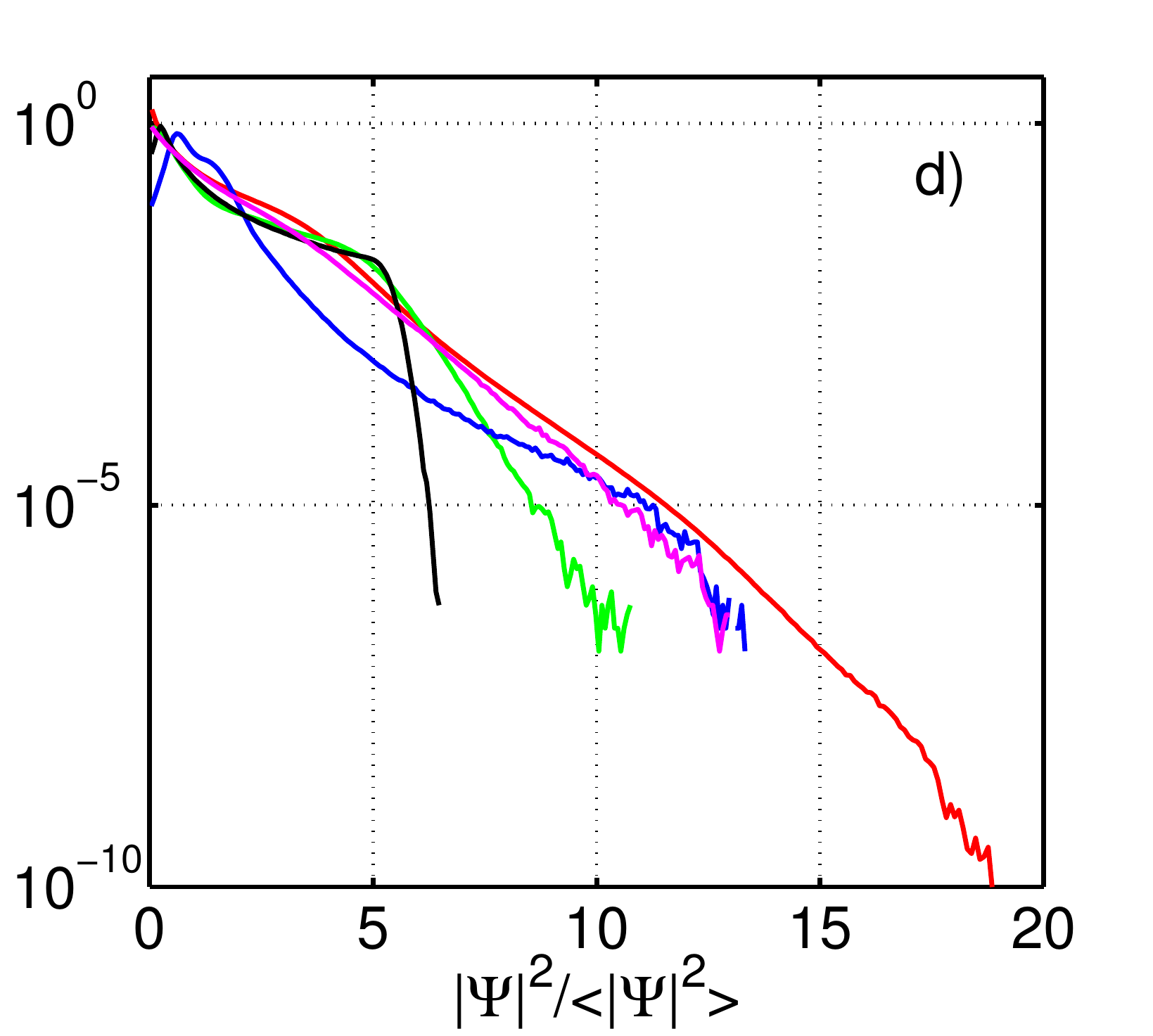}

\caption{\small {\it  (Color on-line) Averaged over ensemble normalized spectra $I_{k}/I_{0}$ at full scale (a) and enlarged at $k=0$ region (b), normalized spacial correlation functions $g(x/x_{corr})/g(0)$ (c) and squared amplitude PDFs (d) for Eq. (\ref{Eq021}) with parameters (\ref{parameters}) at $t=10$ (black), $t=12$ (blue), $t=14$ (green), $t=25$ (purple) and in the statistically state $t\in[200, 250]$ (red). On graph (c) dashed line is Gaussian distribution (\ref{correlation_universal}) and spacial correlation function at $t=12$ was renormalized to fit Gaussian distribution in width at 0.8 level of its maximum.}}
\label{fig:SCPevolution}
\end{figure}

Evolution of spectra, spacial correlation functions and squared amplitude PDFs is shown on FIG.~\ref{fig:SCPevolution}. For not very large time shifts $t<25$ spectra has peak occupying zeroth harmonic $k=0$ only (FIG.~\ref{fig:SCPevolution}b); this peak is clearly seen even during the nonlinear stage of modulation instability development and with high fluctuations in its magnitude it finally vanishes as the system reaches the statistically steady state. In the integrable case (\ref{Eq01}) the peak at zeroth harmonic is always present, while in the nonintegrable NLS equation accounting for small dumping and pumping terms it also gradually vanishes with time \cite{Agafontsev2}. At high wavenumbers $k\gg 1$ spectra approaches to its final shape already to $t=12$ (FIG.~\ref{fig:SCPevolution}a). 

Spacial correlation functions $g(x)$ are connected to spectra $I_{k}$ as
$$
g(x)=\langle\Psi(y,t)\Psi^{*}(y+x,t)\rangle=\int_{-\infty}^{+\infty} I_{k}e^{ikx}\frac{dk}{2\pi}.
$$
Therefore, if energy spectrum has peak at zeroth harmonic, then the corresponding spacial correlation function decays to some non-zero level as $|x|\to +\infty$, as clearly seen on FIG.~\ref{fig:SCPevolution}c. The magnitude of this level fluctuates and finally vanishes as the system approaches to the statistically steady state. Nevertheless, for small lengths $x<x_{corr}$ spacial correlation functions approach to their universal form
\begin{equation}\label{correlation_universal}
g(x/x_{corr})/g(0)\approx \exp\bigg(-4\ln 2\frac{x^{2}}{x_{corr}^{2}}\bigg),
\end{equation}
already to $t\sim 14$. Here $x_{corr}$ is the correlation length defined as full width at half maximum of $g(x)$. It is noteworthy that the peak at zeroth harmonic in spectra and the corresponding non-zero level for spacial correlation functions at infinity both disappear approximately the same time $t\sim 25$ as the oscillations for kinetic $\langle H_{d}\rangle$, four- $\langle H_{4}\rangle$ and higher-order interactions $\langle H_{6}\rangle$ energy cease on FIG.~\ref{fig:EnergyEvolution}.

The shape of the PDFs sharply fluctuates with time as shown on FIG.~\ref{fig:SCPevolution}d, approaching to its final stage only to $t\sim 25$. This behavior is significantly different from that for the system with collapsing focusing six-wave interactions \cite{Agafontsev2} where the PDFs almost immediately took the same shape as for the statistically steady state.

So far we studied the evolution of the statistical characteristics of Eq. (\ref{Eq021}) for the modulation instability development - one of the common scenarios for rogue waves emergence. The next part of the publication is devoted to closer examination of the statistically steady state: as our simulations confirmed, it doesn't depend on initial state, therefore statistical characteristics for Eq. (\ref{Eq021}) will approach with time to that at the statistically steady state. If not stated otherwise, below all of the statistical characteristics relate to the statistically steady states and are additionally averaged over time $t\in [200, 250]$.

\begin{figure}[h] \centering
\includegraphics[width=130pt]{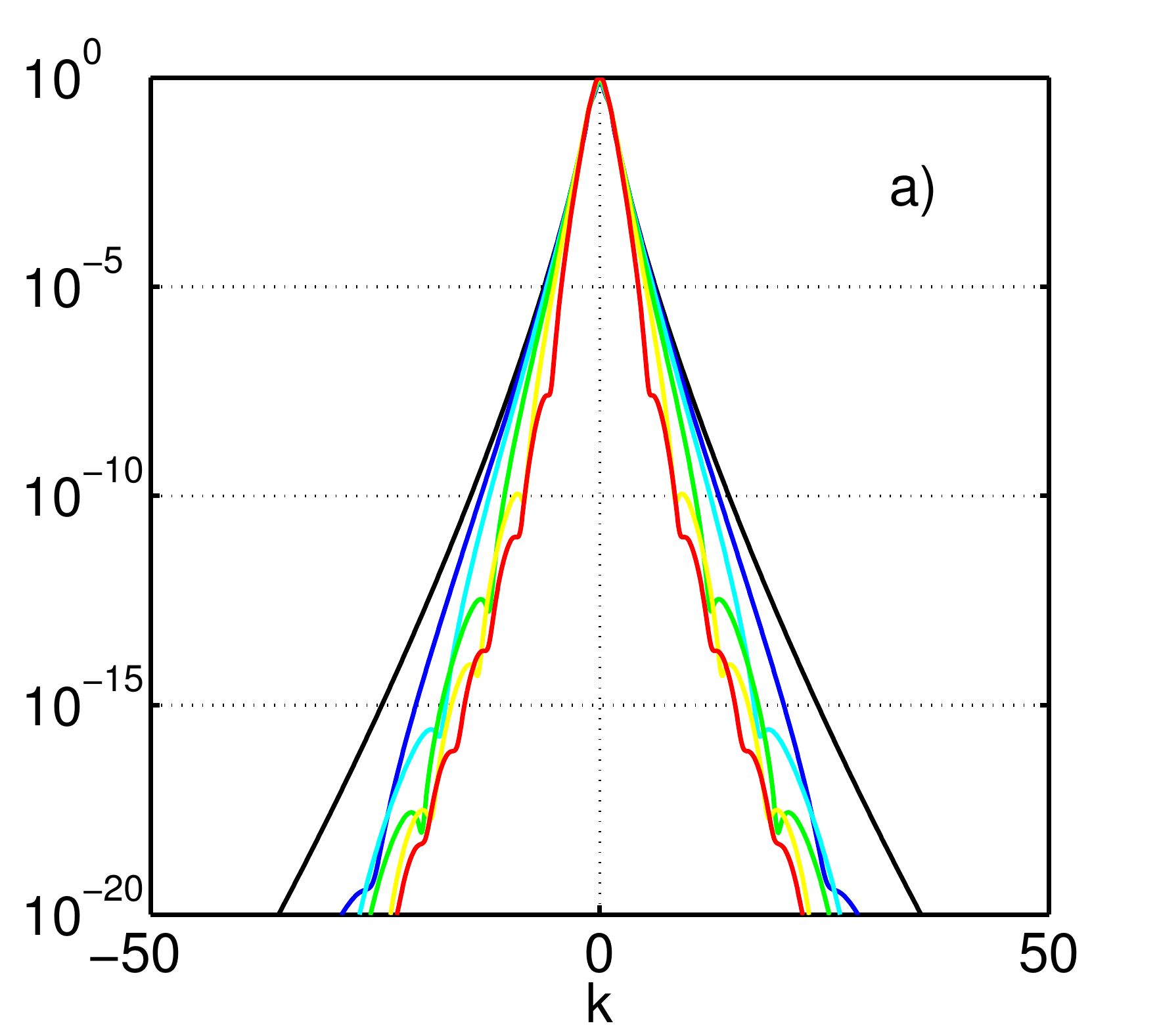}
\includegraphics[width=130pt]{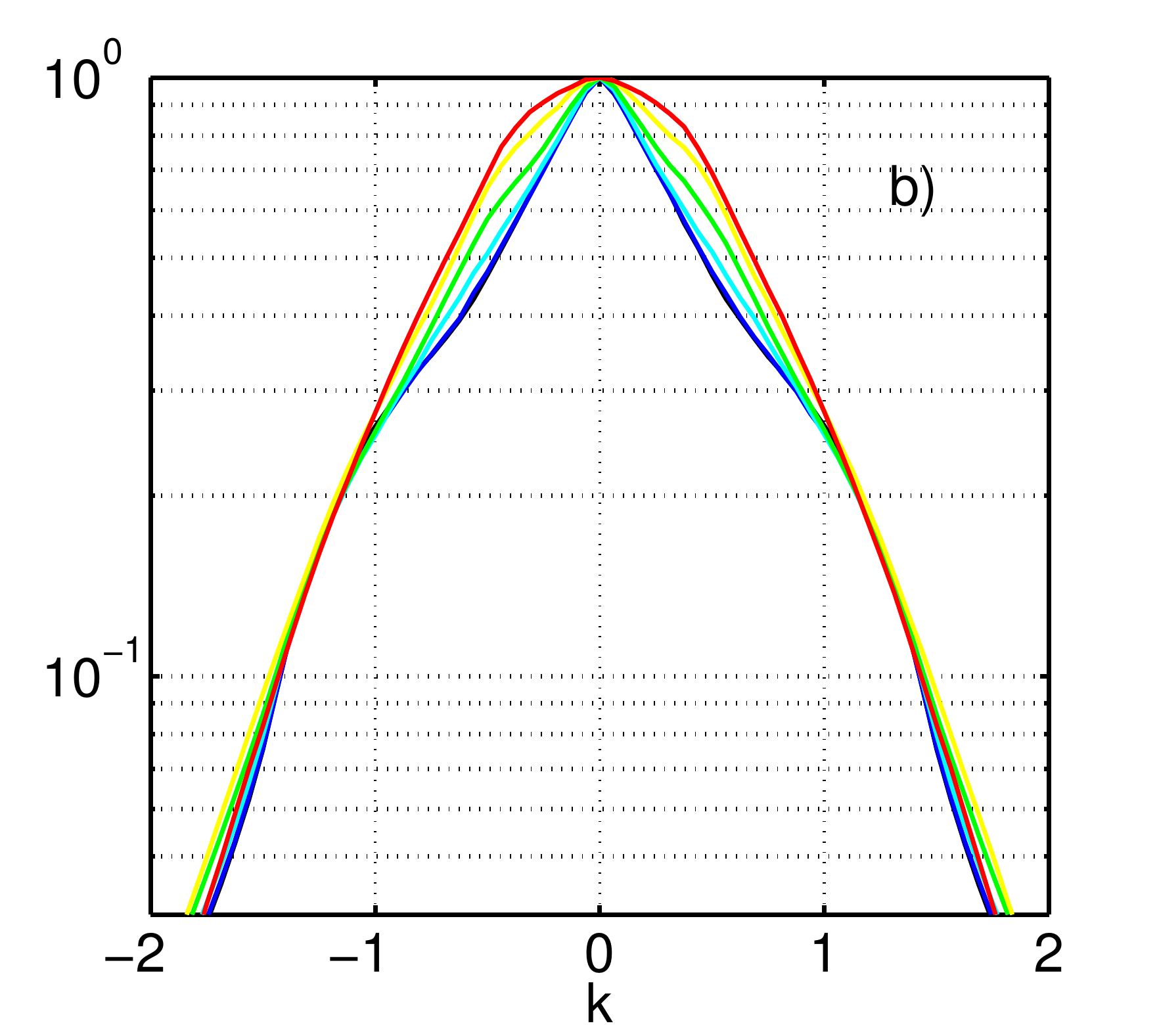}
\includegraphics[width=130pt]{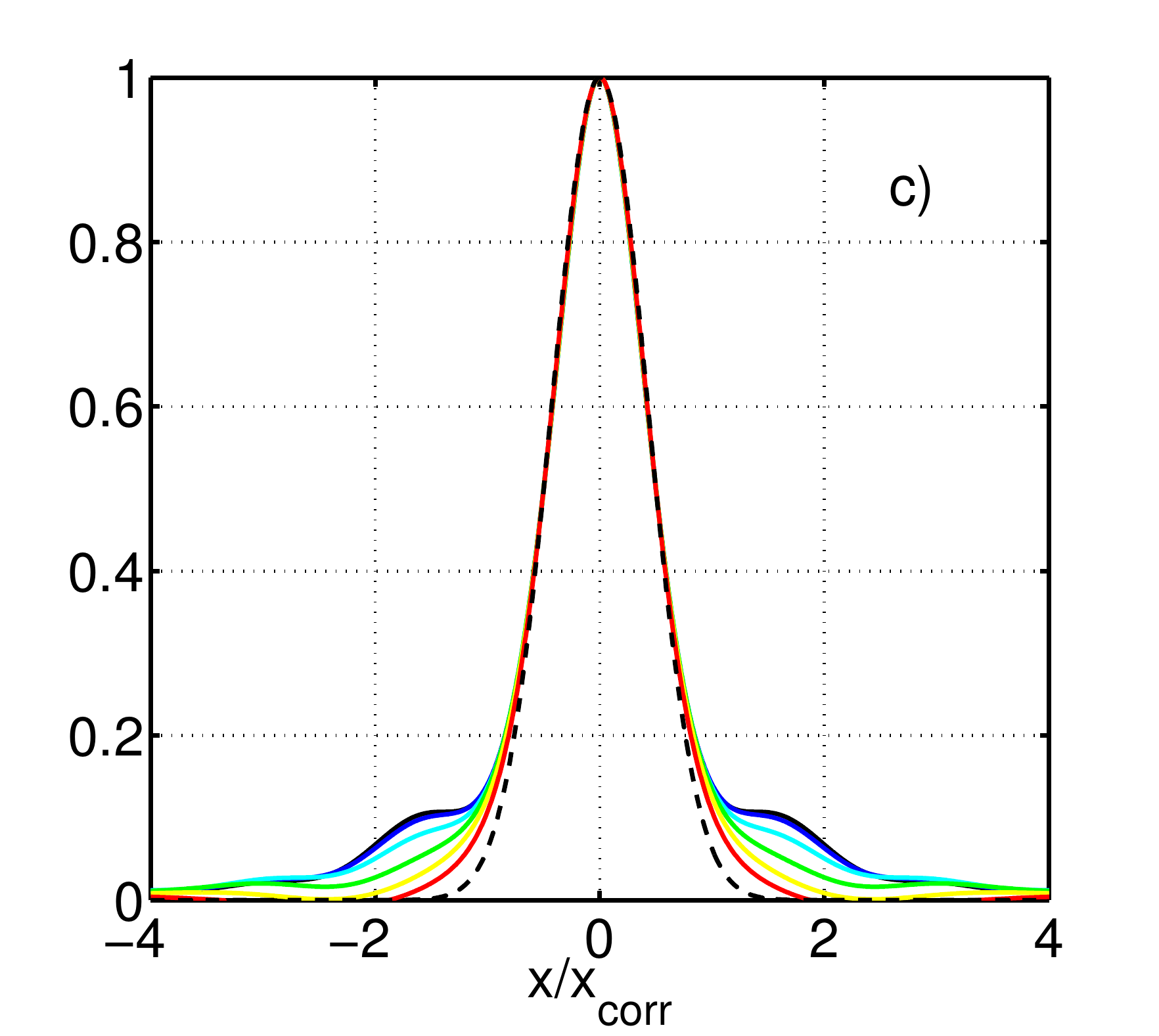}

\caption{\small {\it  (Color on-line) Averaged over ensemble and time $t\in [200, 250]$ normalized spectra $I_{k}/I_{0}$ at full scale (a) and enlarged at $k=0$ region (b) and normalized spatial correlation functions $g(x/x_{corr})/g(0)$ (c) for Eq. (\ref{Eq021}) with $\alpha=0$ (black), $\alpha=0.01$ (blue), $\alpha=0.02$ (cyan), $\alpha=0.04$ (green), $\alpha=0.08$ (yellow), $\alpha=0.16$ (red); $d_{l}=0.0324$, $d_{2p}=0$, $d_{3p}=0.0002$, $p_{1}=0.02$. Dashed black line on graph (c) corresponds to Gaussian distribution (\ref{correlation_universal}).}}
\label{fig:SCalpha}
\end{figure}

\begin{figure}[h] \centering
\includegraphics[width=200pt]{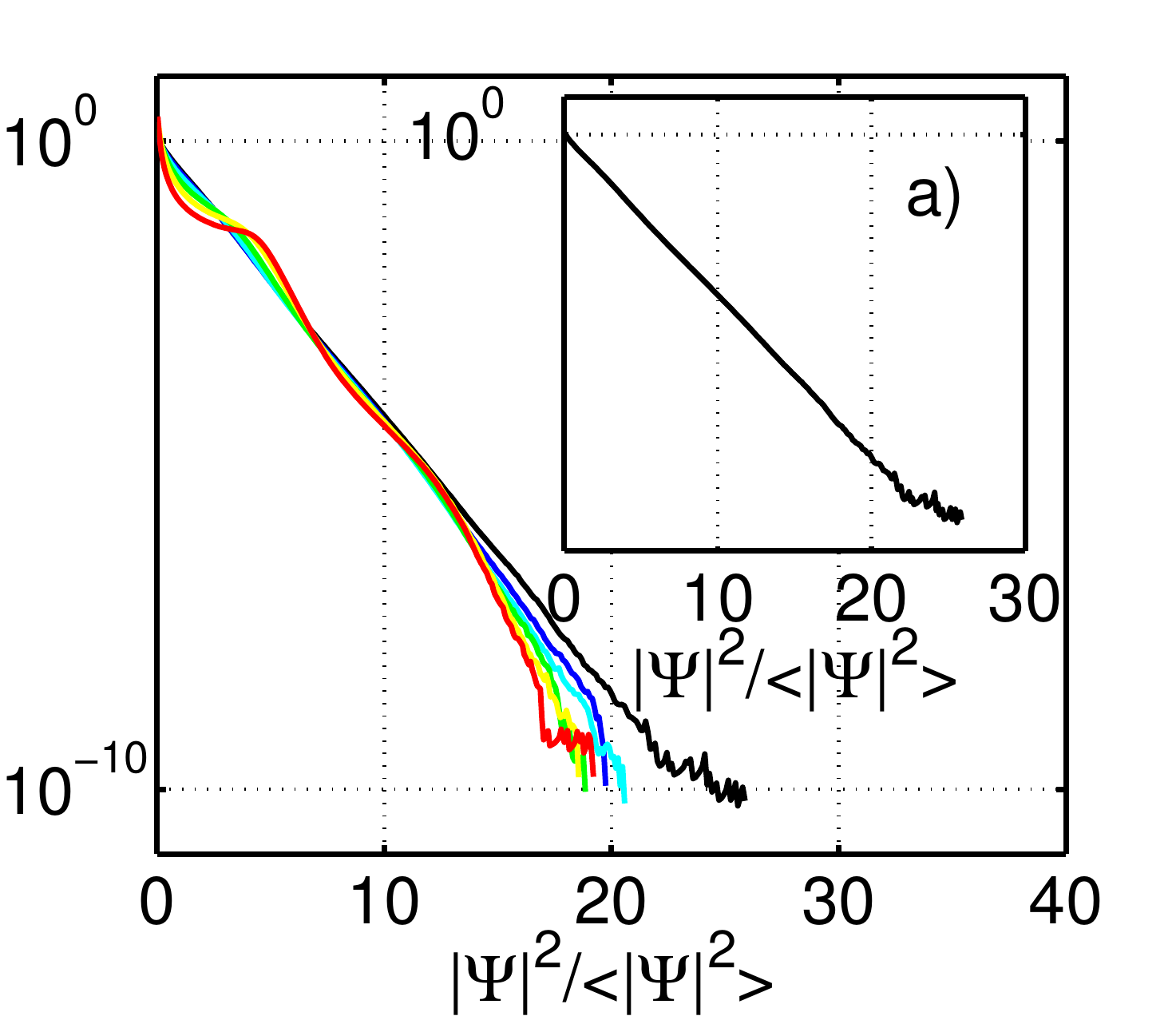}
\includegraphics[width=200pt]{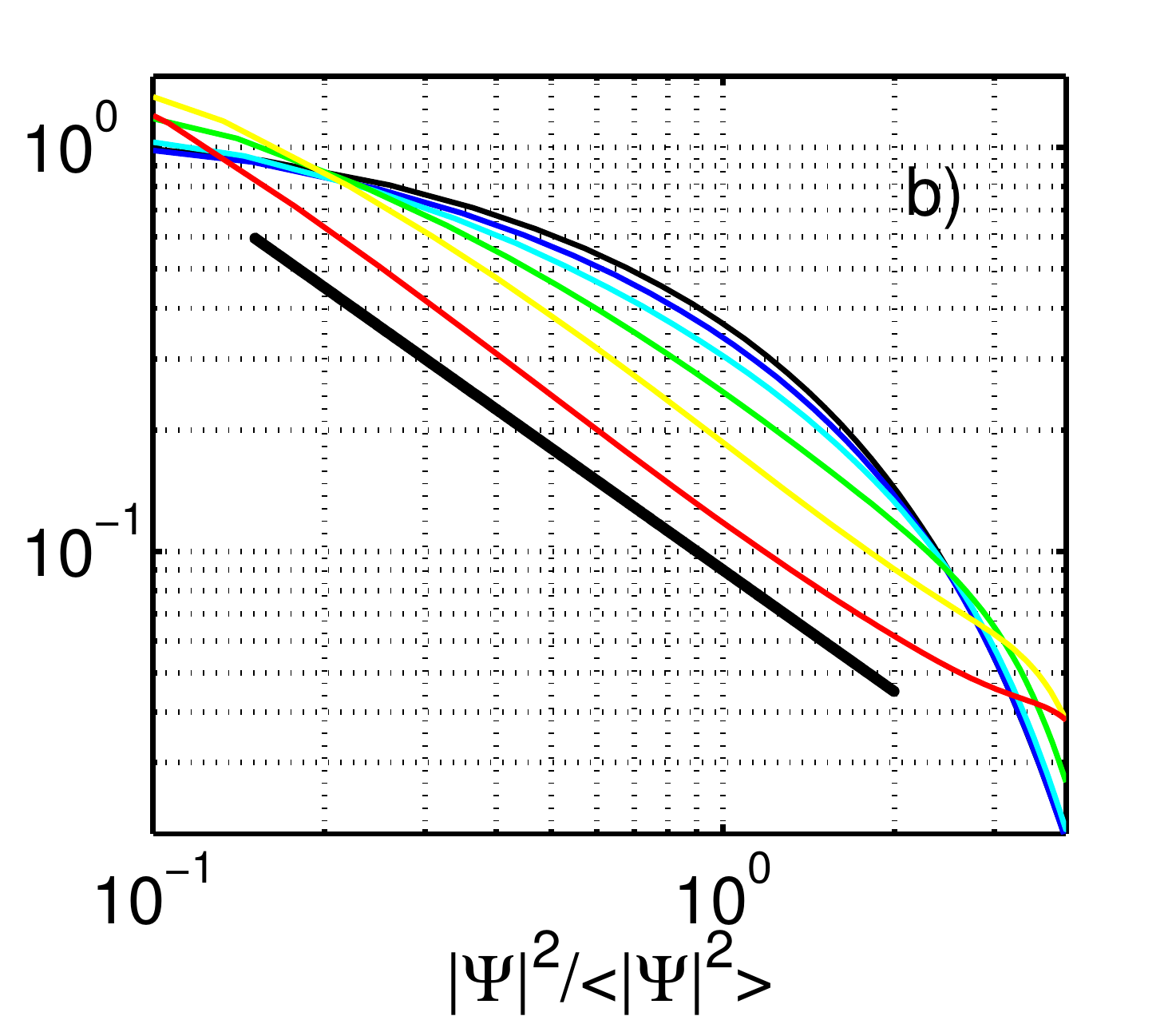}
\includegraphics[width=200pt]{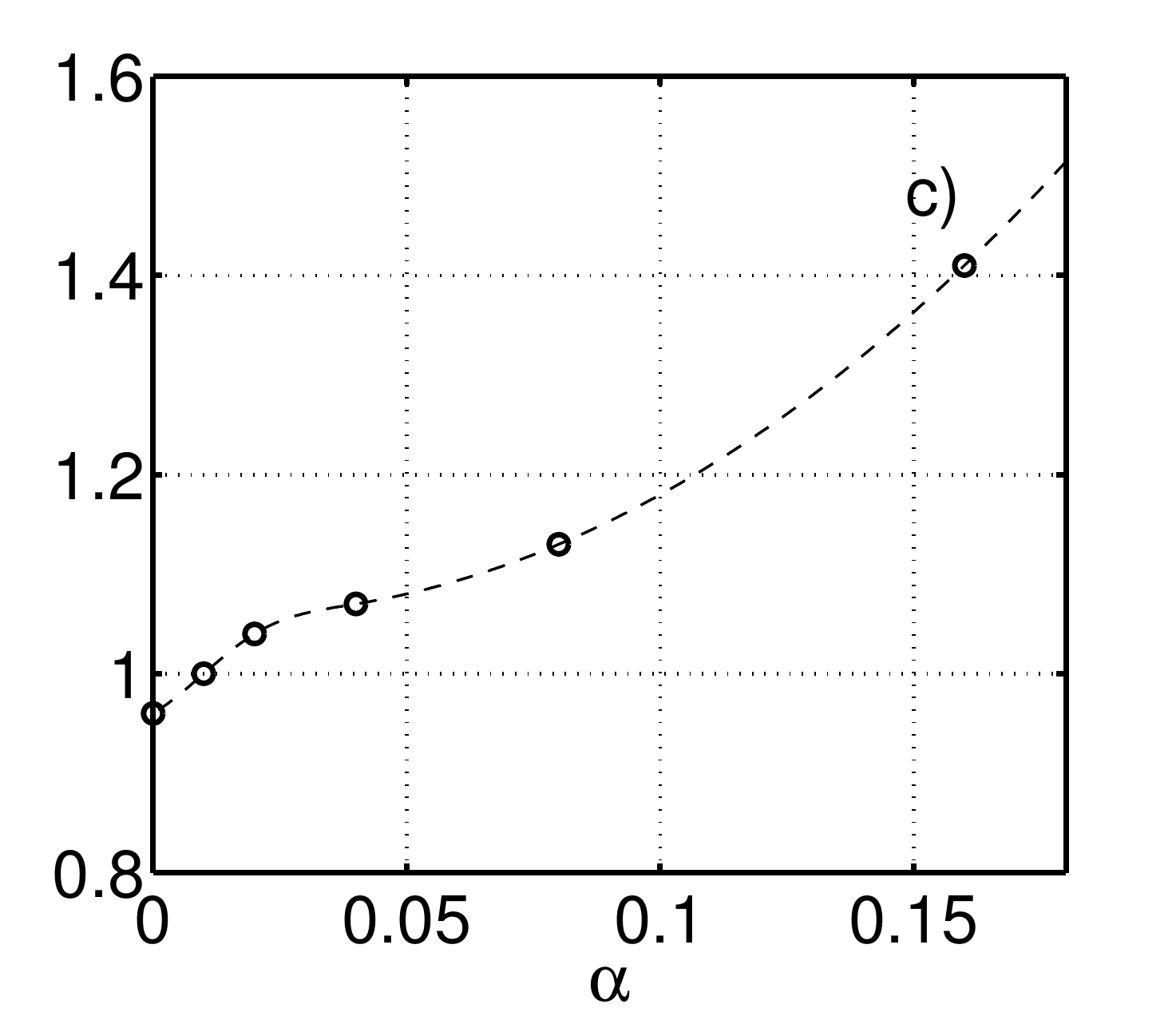}

\caption{\small {\it  (Color on-line) Averaged over ensemble and time $t\in [200, 250]$ squared amplitude PDFs in semi-log scale (a) and enlarged for small and medium amplitudes $|\Psi|^{2}/\langle|\Psi|^{2}\rangle\in[0.1, 4]$ log-log scale (b) and also dependence of mean squared amplitude $\langle|\Psi|^{2}\rangle$ on saturation parameter $\alpha$ (c) for Eq. (\ref{Eq021}) with $\alpha=0$ (black), $\alpha=0.01$ (blue), $\alpha=0.02$ (cyan), $\alpha=0.04$ (green), $\alpha=0.08$ (yellow), $\alpha=0.16$ (red); $d_{l}=0.0324$, $d_{2p}=0$, $d_{3p}=0.0002$, $p_{1}=0.02$. Inset on graph (a) shows squared amplitude PDF for $\alpha=0$ case. Thick line on graph (b) is power-law $\sim x^{-1}$, dashed line on graph (c) is cubic spline fit.}}
\label{fig:Palpha}
\end{figure}

FIG.~\ref{fig:SCalpha} -~\ref{fig:Palpha} demonstrate averaged over ensemble spectra, spacial correlation functions and the PDFs for Eq. (\ref{Eq021}) for six different values of saturation parameter from $\alpha=0$ to $\alpha=0.16$ with all other parameters fixed. We stop at $\alpha=0.16$ because at this point higher-order interactions energy $H_{6}$ first becomes comparable with kinetic energy $H_{d}$. In the absence of saturated nonlinearity $\alpha=0$ averaged spectrum is sharp triangular near the zeroth harmonic $k=0$ and decays monotonically as $k\to +\infty$. This decay is slower near the points $|k|=1$ where maximum growth rate of modulation instability is achieved; on FIG.~\ref{fig:SCalpha}b the corresponding small humps are seen. For high wavenumbers $k\gg\sqrt{2}$ spectrum decays slightly slower than exponentially. In the presence of saturated nonlinearity $\alpha>0$ spectrum shape at $k=0$ becomes smoother, structures near $|k|=1$ become less pronounced, and the spectrum decays faster but non-
monotonically as $k\to +\infty$.

The corresponding normalized spacial correlation functions turn out to be very close to universal Gaussian form (\ref{correlation_universal}) for small lengths $|x|<x_{corr}$. Beyond the correlation length $|x|>x_{corr}$ the correlation functions decay to zero level with small oscillations; the decay is faster for higher saturation parameters $\alpha$.

In the absence of saturated nonlinearity $\alpha=0$ the PDF is purely Rayleigh one as shown on the inset of FIG.~\ref{fig:Palpha}a (see also \cite{Agafontsev2}), even despite the fact that the system is in significantly nonlinear regim $|H_{4}|\sim |H_{d}|$. Here and below we plot most of the graphs for squared amplitude PDFs versus $|\Psi|^{2}/\langle|\Psi|^{2}\rangle$ where $\langle|\Psi|^{2}\rangle$ is the mean over ensemble squared amplitude: mean wave action $\langle N\rangle$ and squared amplitude $\langle|\Psi|^{2}\rangle=\langle N\rangle/\int dx$ depend slightly on saturation parameter $\alpha$ and significantly on dumping and pumping parameters $d_{l}$, $d_{2p}$, $d_{3p}$ and $p_{1}$. 

Addition of saturated nonlinearity significantly modifies the PDFs. The most interesting result here is the power-law region for small $|\Psi|\ll\sqrt{\langle|\Psi|^{2}\rangle}$ and medium $|\Psi|\sim\sqrt{\langle|\Psi|^{2}\rangle}$ amplitudes that is clearly visible starting from saturation parameters $\alpha=0.04$, as shown of FIG.~\ref{fig:Palpha}b. This power-law region is absent for $\alpha<0.04$, and for larger saturation parameters it extends itself into regions of medium and especially small amplitudes with increase of $\alpha$. All of the lines shown of FIG.~\ref{fig:Palpha}b have almost the same slope at $|\Psi|^{2}/\langle|\Psi|^{2}\rangle= 1$: from -1.07 to -0.92. The latter means that the PDFs in this region decay as 
\begin{equation}\label{Eq7}
PDF(|\Psi|^{2}/\langle|\Psi|^{2}\rangle)\sim \bigg(|\Psi|^{2}/\langle|\Psi|^{2}\rangle\bigg)^{-1}.
\end{equation}
This is an amazing result especially taking into account that saturated nonlinearity becomes important for systems where extremely high amplitudes are achieved. Indeed, average waves $|\Psi|^{2}\sim\langle|\Psi|^{2}\rangle$ that occur in the system most frequently and that can be very high in the physical variables (but are still small in the sense of saturated nonlinearity $\alpha|\Psi|^{2}\ll 1$) turn out to be distributed by power-law (\ref{Eq7}) or in terms of amplitude PDFs as $PDF(|\Psi|)\sim|\Psi|^{-1}$, so that waves with twice different amplitudes occur with only two times different frequencies.

The power-law region (\ref{Eq7}) for small and medium waves is followed by intermediate region for larger amplitudes and then by Rayleigh tail for high amplitudes. The latter result is straightforward: for very high waves $\alpha|\Psi|^{2}\gg 1$ saturated nonlinearity term becomes linear one, 
$$
\frac{|\Psi|^2}{1+\alpha|\Psi|^2}\Psi\approx \frac{\Psi}{\alpha},
$$
so that Eq. (\ref{Eq021}) contains large linear and small nonlinear terms.

For small saturation parameters $\alpha$ mean squared amplitude $\langle|\Psi|^{2}\rangle$ increases almost linearly with $\alpha$, as demonstrated on FIG.~\ref{fig:Palpha}c. The level of growth drops substantially at $\alpha\sim 0.02-0.03$ that might indicate in favor of some changes in the interior processes of the system. For larger saturation parameters the growth continues and gradually accelerates with $\alpha$. It is interesting to note that the power-law region first becomes visible starting approximately from $\alpha=0.04$.

\begin{figure}[h] \centering
\includegraphics[width=200pt]{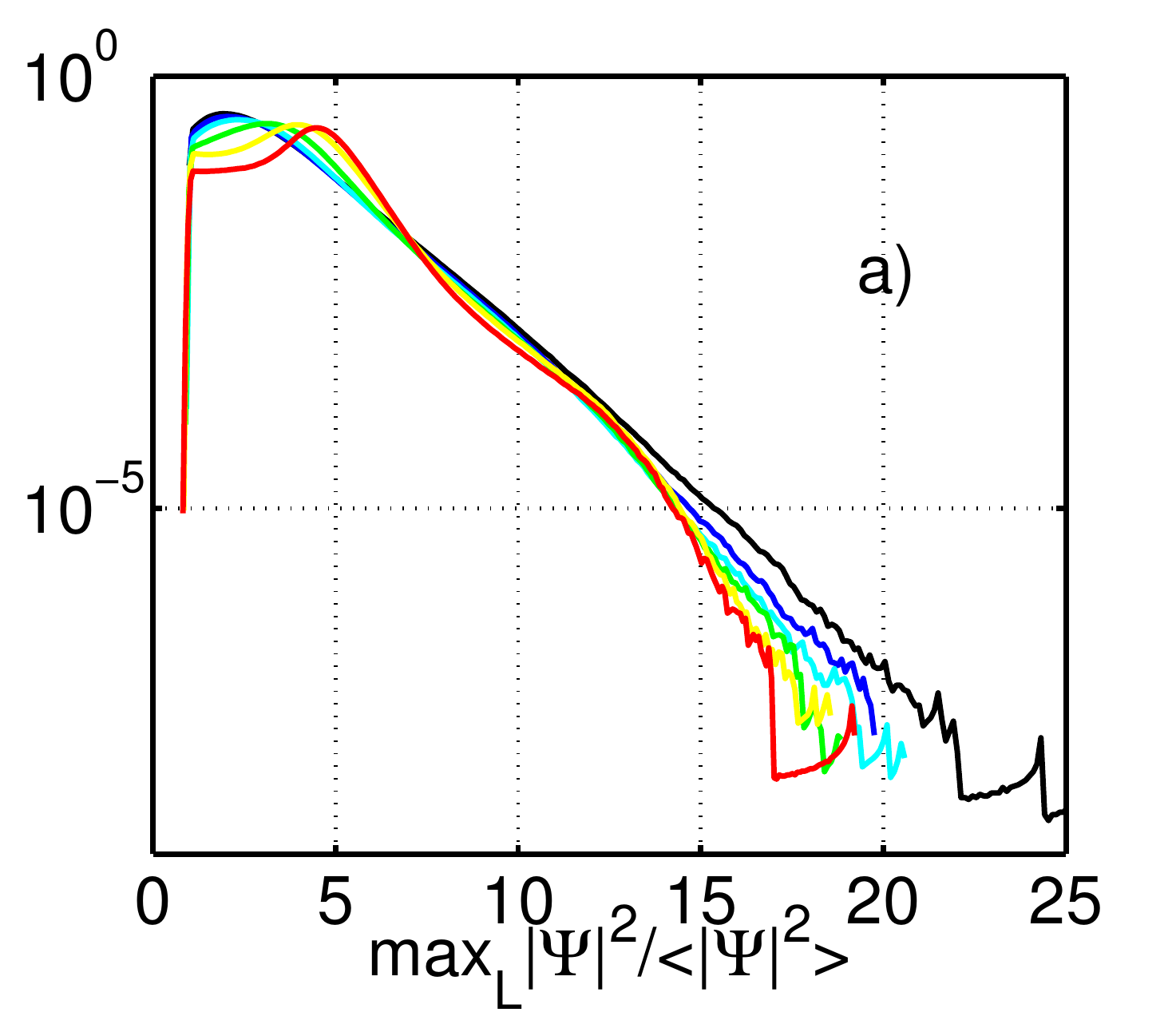}
\includegraphics[width=200pt]{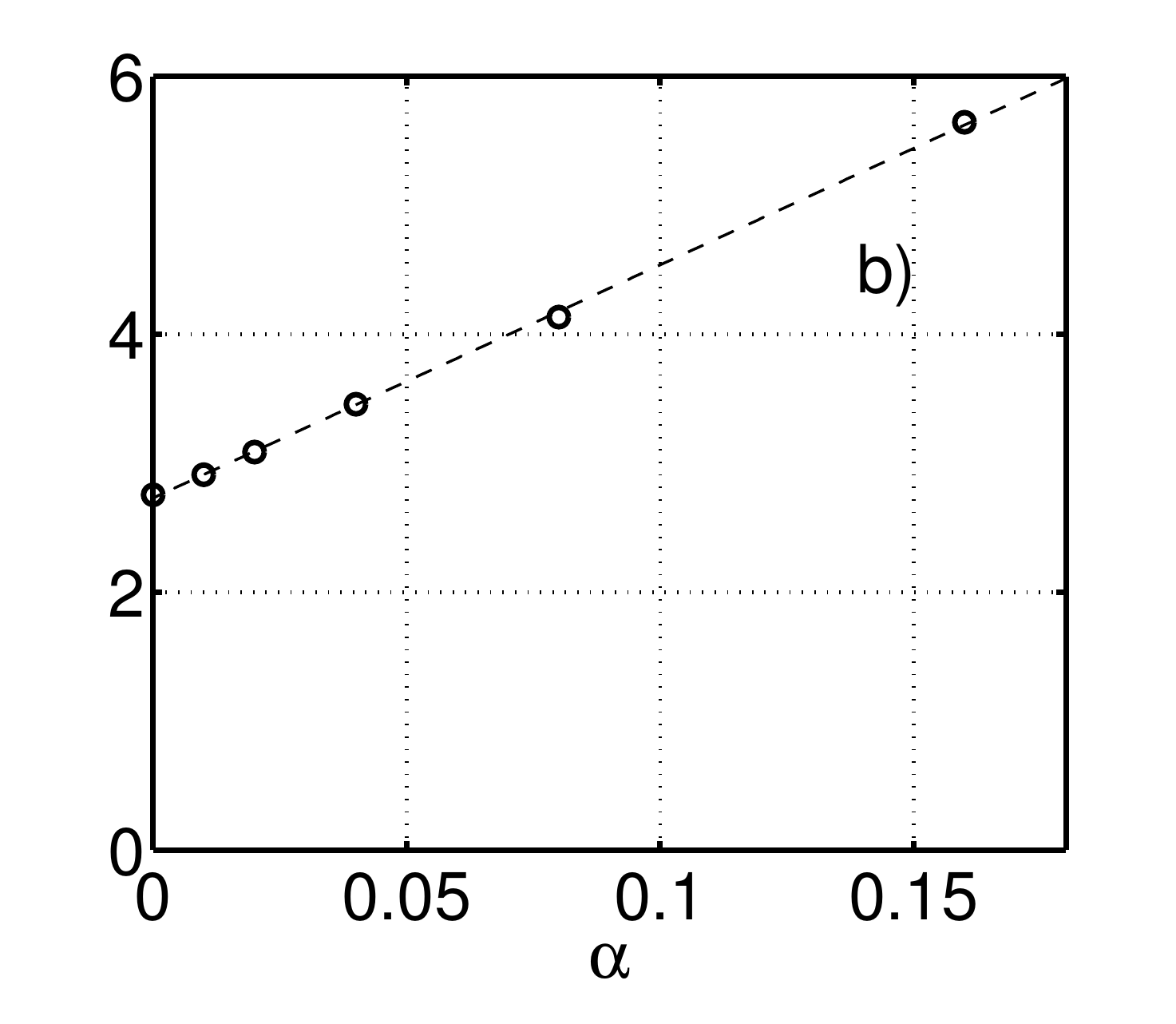}

\caption{\small {\it  (Color on-line) Averaged over ensemble and time $t\in [200, 250]$ PDFs for squared local maximums of $|\Psi|$ (a) and dependence of mean squared local maximum amplitude $\langle\max_{L}|\Psi|^{2}\rangle$ on saturation parameter $\alpha$ (b) for Eq. (\ref{Eq021}) with $\alpha=0$ (black), $\alpha=0.01$ (blue), $\alpha=0.02$ (cyan), $\alpha=0.04$ (green), $\alpha=0.08$ (yellow), $\alpha=0.16$ (red); $d_{l}=0.0324$, $d_{2p}=0$, $d_{3p}=0.0002$, $p_{1}=0.02$. Dashed line on graph (b) is linear fit.}}
\label{fig:Plocal_alpha}
\end{figure}

In addition to PDFs for amplitudes of the entire field $\Psi$ we also measured PDFs for local maximums of $|\Psi|$. In order to make such PDFs more physically relevant in the sense of what an external observer would see, we filtered out sufficiently small local maximums $(\max_{L}|\Psi|^{2})/\langle|\Psi|^{2}\rangle<1$ (see the corresponding drop at this threshold on FIG.~\ref{fig:Plocal_alpha}a) and also such local maximums that represented the same wave in the reality. In particular, according to our measurements technique, several local maximums on a hump of a large wave (a large wave perturbed by small high-frequency modulation) contributed to the PDFs as only one - the highest - local maximum. 

As shown of FIG.~\ref{fig:Plocal_alpha}a, the PDFs for squared local maximums are very similar to the PDFs for entire field $|\Psi|^{2}$: humps in the region of medium amplitudes on FIG.~\ref{fig:Plocal_alpha}a are situated at the same points as humps in the end of power-law regions on FIG.~\ref{fig:Palpha}a, then both types of the PDFs have very similar intermediate regions for higher waves followed by Rayleigh far tails. It is noteworthy that average squared local maximum amplitude $\langle\max_{L}|\Psi|^{2}\rangle$ increases with saturation parameter $\alpha$ almost linearly (FIG.~\ref{fig:Plocal_alpha}b), contrary to more complex behavior of the mean squared amplitude $\langle|\Psi|^{2}\rangle$ for entire field, shown on FIG.~\ref{fig:Palpha}c. Note that $\langle\max_{L}|\Psi|^{2}\rangle$ was measured using filtered set of local maximums.

\begin{figure}[h] \centering
\includegraphics[width=200pt]{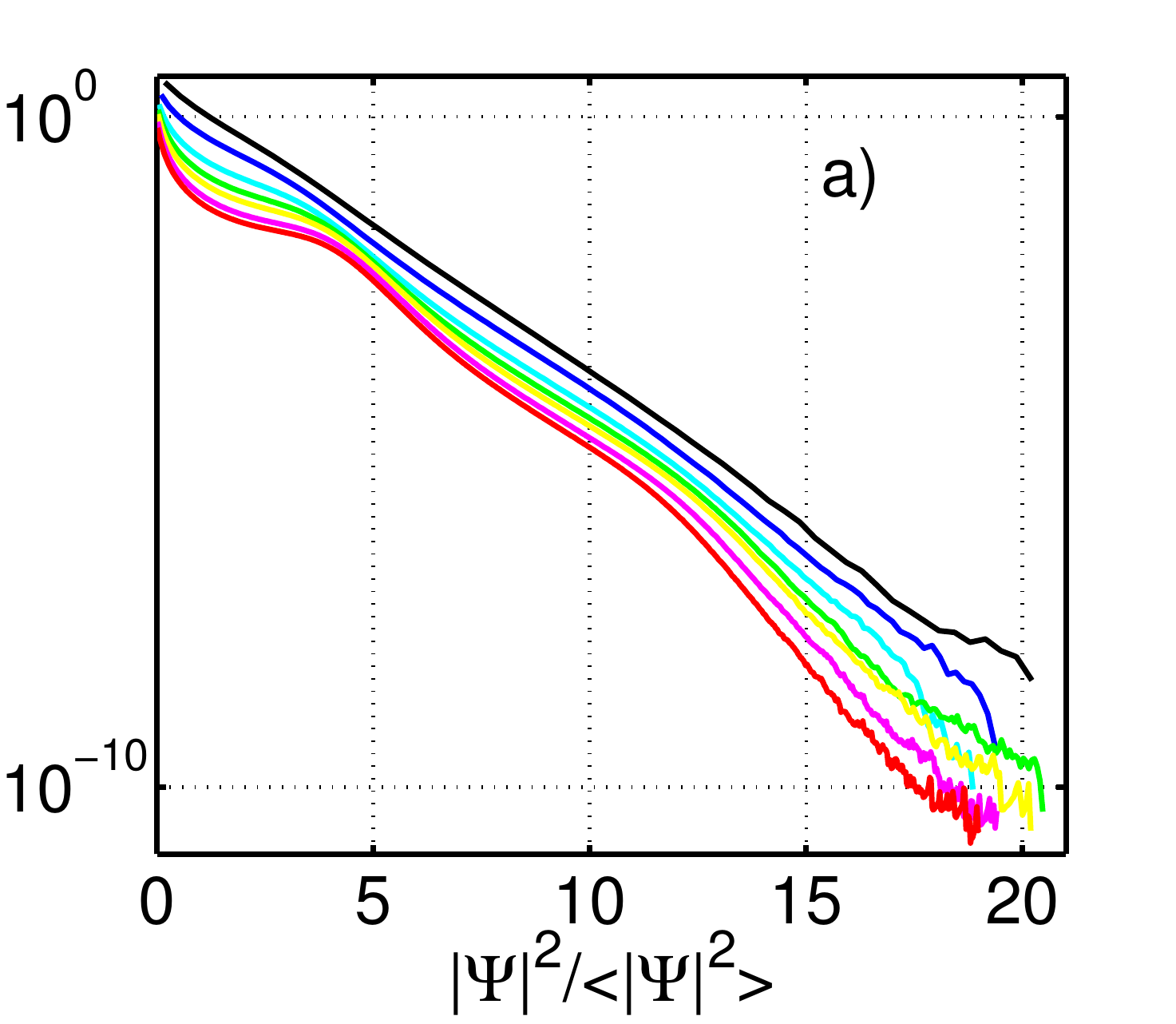}
\includegraphics[width=200pt]{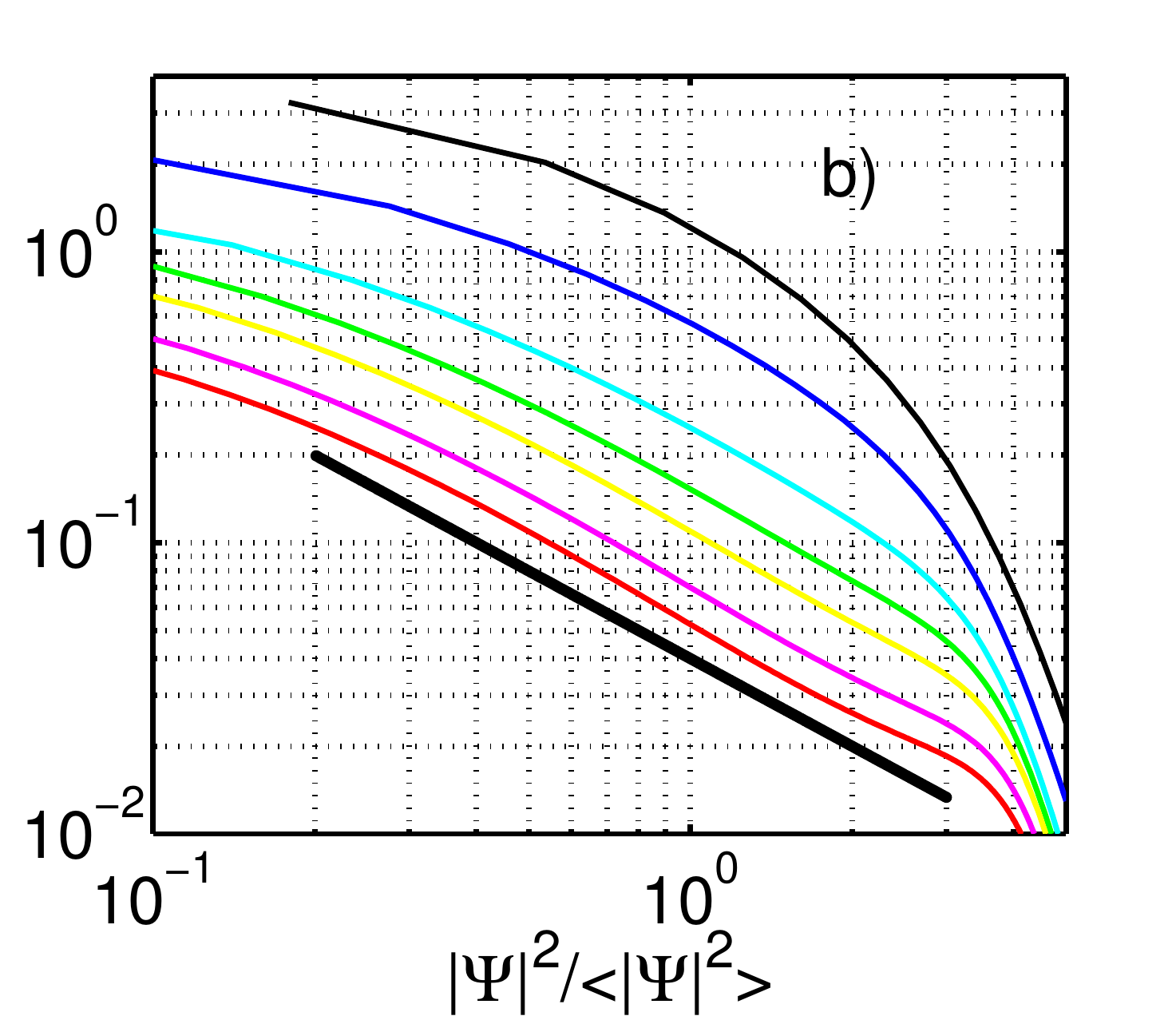}
\includegraphics[width=200pt]{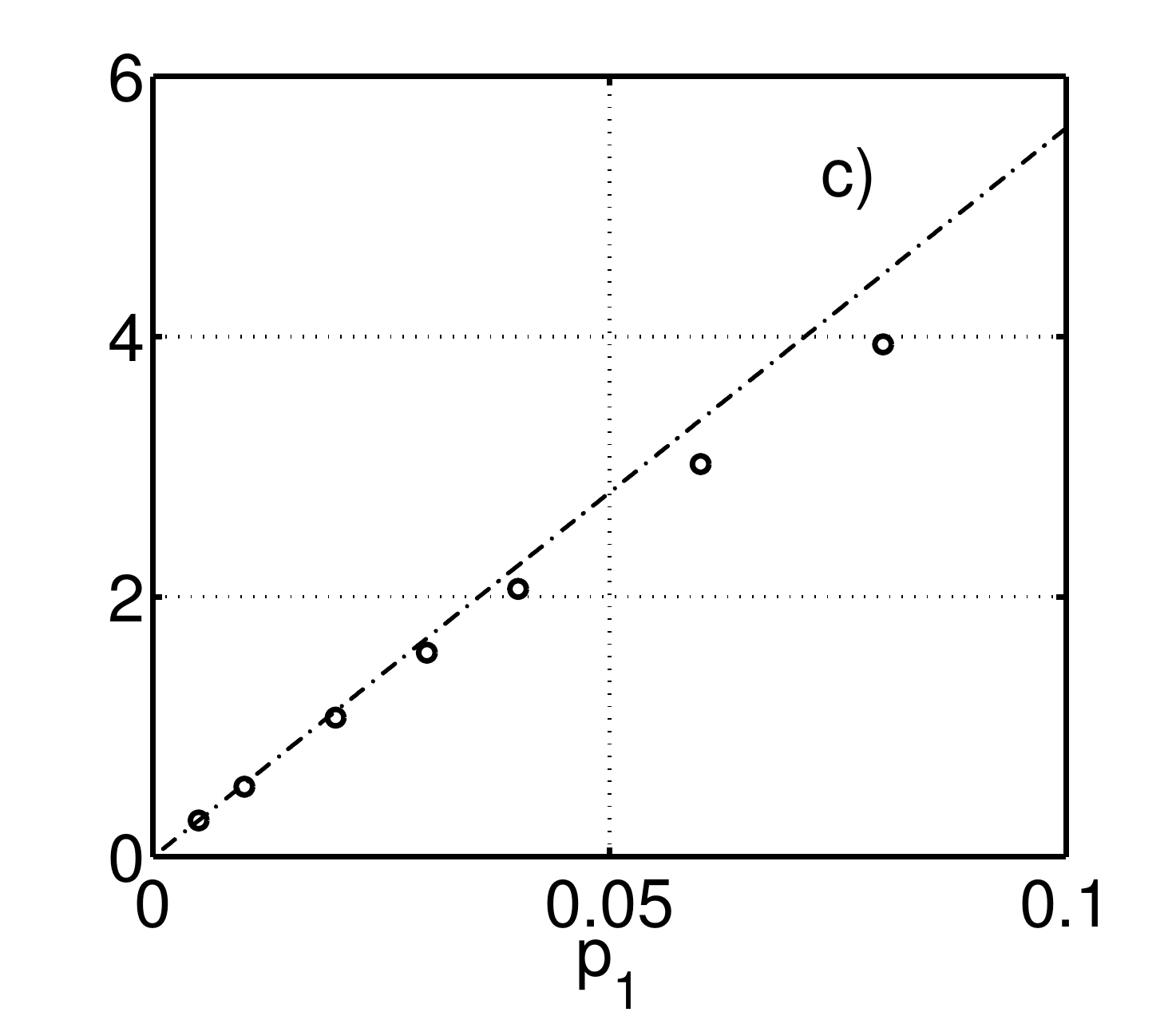}

\caption{\small {\it  (Color on-line) Averaged over ensemble and time $t\in [200, 250]$ squared amplitude PDFs in semi-log scale (a) and enlarged for small and medium amplitudes $|\Psi|^{2}/\langle|\Psi|^{2}\rangle\in[0.1, 5]$ log-log scale (b) and also dependence of mean squared amplitude $\langle|\Psi|^{2}\rangle$ on pumping coefficient $p_{1}$ (c) for Eq. (\ref{Eq021}) with $p_{1}=0.005$ (black), $p_{1}=0.01$ (blue), $p_{1}=0.02$ (cyan), $p_{1}=0.03$ (green), $p_{1}=0.04$ (yellow), $p_{1}=0.06$ (purple), $p_{1}=0.08$ (red); $\alpha=0.04$, $d_{l}=0.0324$, $d_{2p}=0$, $d_{3p}=0.0002$. Thick line on graph (b) is power-law $\sim x^{-1}$, dashed line on graph (c) is straight tangent line to the dependence of mean squared amplitude at small $p_{1}$.}}
\label{fig:Pp1}
\end{figure}

FIG.~\ref{fig:Pp1} demonstrates the PDFs for Eq. (\ref{Eq021}) with fixed saturation parameter $\alpha=0.04$ and dumping coefficients $d_{l}=0.0324$, $d_{2p}=0$, $d_{3p}=0.0002$, while the pumping coefficient varies from $p_{1}=0.005$ to $p_{1}=0.08$. Dumping and pumping coefficients determine the statistically steady state, so that the average wave amplitude $C=\sqrt{\langle|\Psi|^{2}\rangle}$ in this state increases with $p_{1}$ (see FIG.~\ref{fig:Pp1}c) almost as 
$$
C\sim\sqrt{p_{1}},\quad p_{1}\ll 1.
$$
With the help of the scaling and gauge transformations $\Psi=C\tilde{\Psi}e^{i(1-1/C^{2})\tilde{t}}$, $x=\tilde{x}/C$ and $t=\tilde{t}/C^{2}$, Eq. (\ref{Eq021}) with deterministic pumping term $\Phi=p_{1}\Psi$ is rewritten as
\begin{equation}\label{Eq0211}
i\Psi_t - \Psi + (1-id_{l})\Psi_{xx}+\frac{|\Psi|^2}{1+\alpha C^{2}|\Psi|^2}\Psi + id_{2p}|\Psi|^2\Psi + id_{3p}C^{2}|\Psi|^4\Psi = i\frac{p_{1}}{C^{2}}\Psi,
\end{equation}
where all tilde signs are omitted. Therefore, magnification of the pumping coefficient is equivalent to increase of saturation parameter $\alpha$ with modified pumping and dumping coefficients $d_{3p}$ and $p_{1}$. In particular, for sufficiently small dumping and pumping a set of coefficients $(\alpha, d_{l}, d_{2p}, d_{3p}, \kappa p_{1})$ is equivalent to $(\kappa\alpha, d_{l}, d_{2p}, \kappa d_{3p}, p_{1})$ if $\kappa$ is not very different from 1. In this sense the results shown on FIG.~\ref{fig:Pp1}a,b have direct correspondence with that on FIG.~\ref{fig:Palpha}a,b: the PDFs for small or large pumping coefficients $p_{1}$ are very similar to the PDFs for small or large saturation parameters $\alpha$ respectively. Power-law region is absent for small pumping coefficients up to $p_{1}\sim 0.02$ and then becomes more pronounced with $p_{1}$. The same conclusions are for modifications of the dumping parameters $d_{l}$, $d_{2p}$ and $d_{3p}$: their magnification leads to decrease of the average amplitude in 
the statistically steady state $C$ and therefore is equivalent to decrease of saturation parameter $\alpha$ with modified pumping and dumping coefficients.

\begin{figure}[h] \centering
\includegraphics[width=200pt]{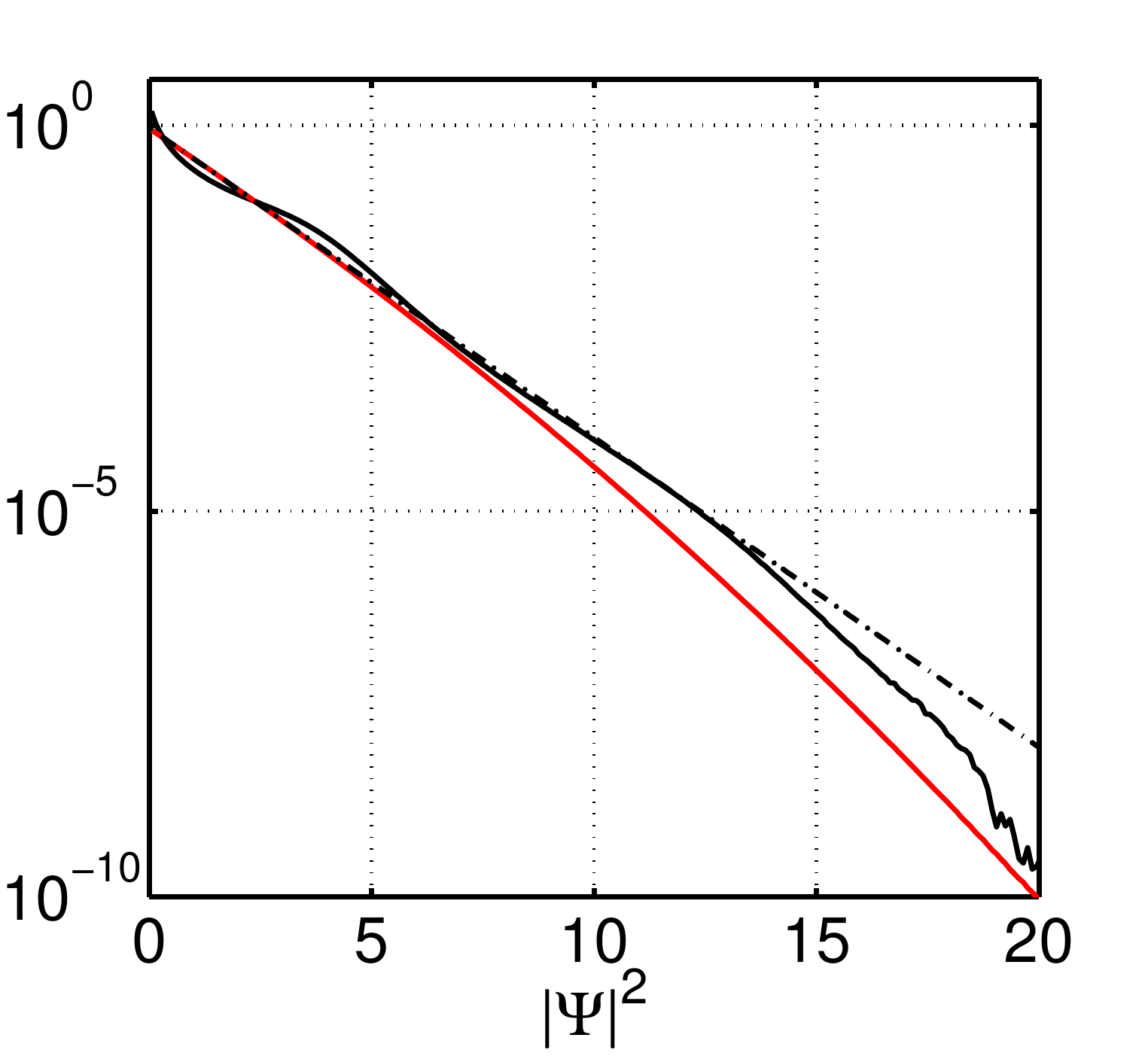}

\caption{\small {\it  (Color on-line) Averaged over ensemble and time $t\in [200, 250]$ squared amplitude PDFs for pure solution of Eq. (\ref{Eq021}) $\Psi$ (black) and modified solution of Eq. (\ref{Eq021}) $\tilde{\Psi}=\mathscr{F}^{-1}(\mathscr{F}(\Psi)\times e^{i\xi_{k}})$ shifted by arbitrary phases $\xi_{k}$ (red) where $\mathscr{F}$ is Fourier transform; values for coefficients are the same as in (\ref{parameters}). Black dashed line is tangent straight line to the PDF of modified solutions $PDF(|\tilde{\Psi}|^{2})$.}}
\label{fig:PLSH}
\end{figure}

At fixed time $t$ a given state of wave field $\Psi$ can be expanded in series of linear waves (\ref{DFT}). The amplitudes of linear waves are determined through energy spectrum as $a_{k}(t)=|\Psi_{k}(t)|$. Correlation of waves phases $\phi_{k}$ is usually very important for the generation of extreme events. FIG.~\ref{fig:PLSH} shows two PDFs: one (black curve) for numerical solutions $\Psi$ of Eq. (\ref{Eq021}), and the other (red curve) for wave fields $\tilde{\Psi}$ obtained from $\Psi$ by shifting phases by arbitrary values. Both fields $\Psi$ and $\tilde{\Psi}$ have the same spectra and spacial correlation functions.

Technically the second PDF was calculated the same way as the first one, but instead of one numerical solution $\Psi(x,t)$ for the given time step and initial condition $\Psi(t=0)$ we took several (usually $M=50$) wave fields $\tilde{\Psi}=\mathscr{F}^{-1}(\mathscr{F}(\Psi)\times e^{i\xi_{k}})$ for several different realizations of arbitrary phases $\xi_{k}$. Here $\mathscr{F}$ is Fourier transform. We checked that number $M=50$ we used was sufficient for solid results and its magnification did not lead to any further changes in the resulting PDF. So, by its construction the second PDF is composed of the same variants of amplitudes $a_{k}=|\Psi_{k}|$ and the only difference from the PDF of Eq. (\ref{Eq021}) is the detuning of phases.

When phases are detuned, the corresponding PDF turns out to be very close to Rayleigh one but decays slightly faster. Events with high amplitudes occur about one order of magnitude less frequently in this case starting from squared amplitudes $|\Psi|^{2}>10$. Nevertheless, the influence of correlation of phases is much less significant overall for Eq. (\ref{Eq021}) than in case of collapsing focusing six-wave interactions studied in \cite{Agafontsev2}.

\begin{figure}[h] \centering
\includegraphics[width=130pt]{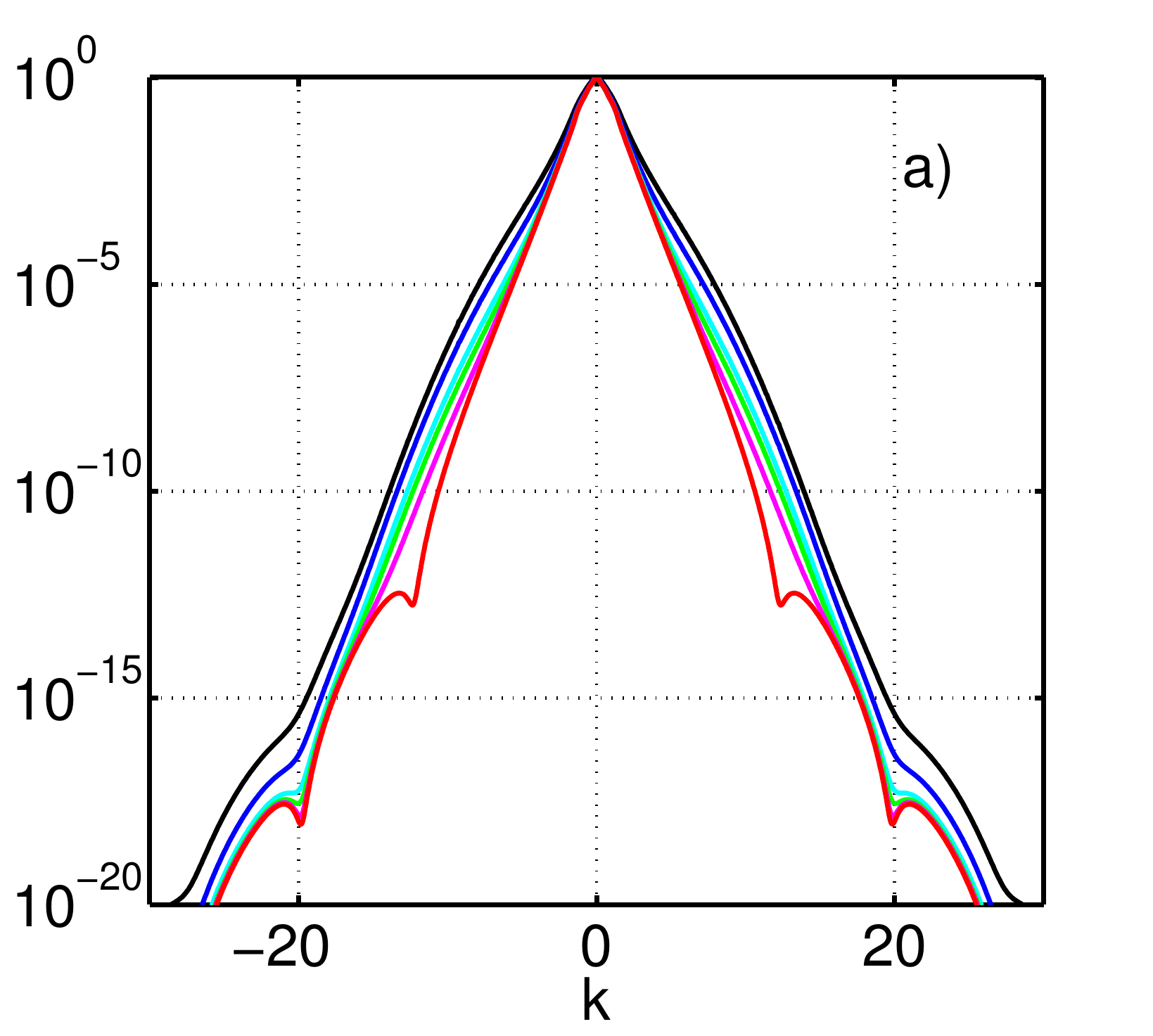}
\includegraphics[width=130pt]{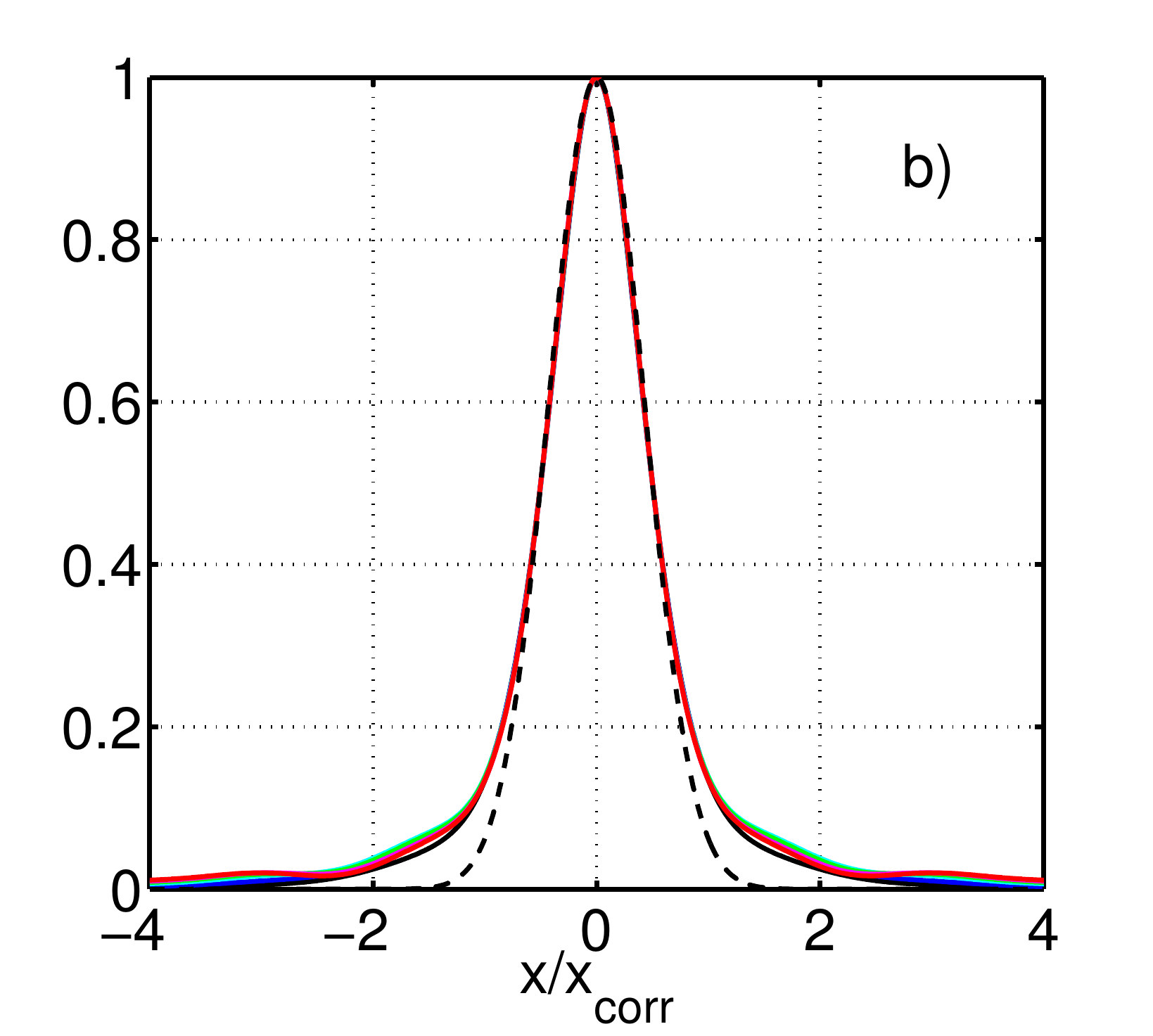}
\includegraphics[width=130pt]{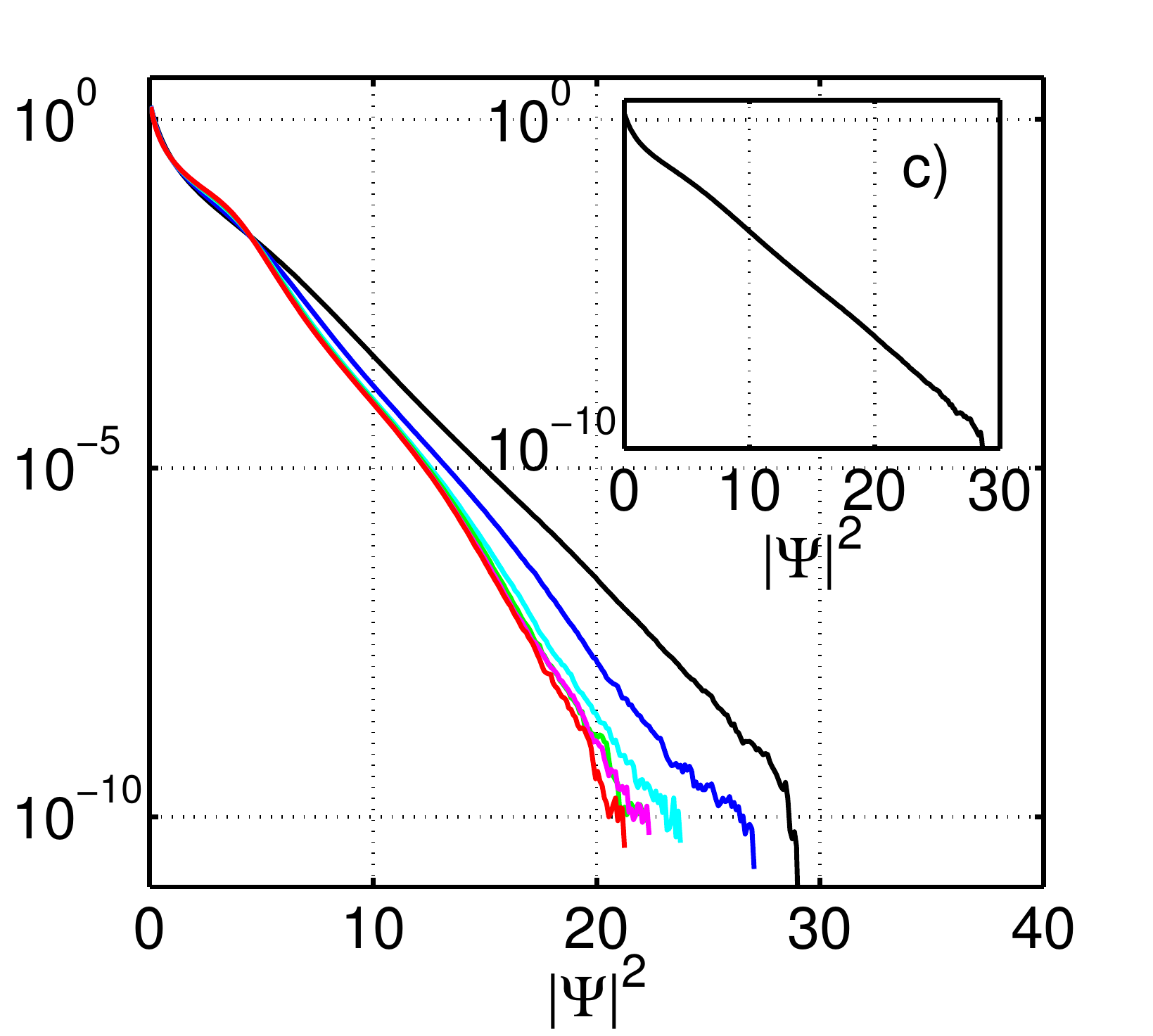}

\caption{\small {\it  (Color on-line) Averaged over ensemble and time $t\in [200, 250]$ normalized spectra $I_{k}/I_{0}$ (a), normalized spacial correlation functions $g(x/x_{corr})/g(0)$ (b) and squared amplitude PDFs (c) for Eq. (\ref{Eq021}) with superimposed deterministic and stochastic forcing terms $\Phi=p_{1}\Psi+\Phi_{2}$ for stochastic pumping term coefficients $p_{2}=100$ (corresponds to average stochastic pumping amplitude $\sqrt{\langle|\Phi_{2}|^{2}\rangle_{\xi}}\approx 10$, black curve), $p_{2}=60$ ($\sqrt{\langle|\Phi_{2}|^{2}\rangle_{\xi}}\approx 6$, blue), $p_{2}=30$ ($\sqrt{\langle|\Phi_{2}|^{2}\rangle_{\xi}}\approx 3$, cyan), $p_{2}=20$ ($\sqrt{\langle|\Phi_{2}|^{2}\rangle_{\xi}}\approx 2$, green), $p_{2}=10$ ($\sqrt{\langle|\Phi_{2}|^{2}\rangle_{\xi}}\approx 1$, purple), $p_{2}=10^{-4}$ ($\sqrt{\langle|\Phi_{2}|^{2}\rangle_{\xi}}\approx 10^{-5}$, red); $\theta_{p}=5$, values for other coefficients are the same as in (\ref{parameters}). Dashed line on graph (b) shows Gaussian 
distribution (\ref{correlation_universal}), inset on graph (c) shows the PDF for $p_{2}=100$ case.}}
\label{fig:stochastic}
\end{figure}

So far we examined Eq. (\ref{Eq021}) with deterministic pumping term (or positive feedback), while for real systems stochastic forcing may be important. FIG.~\ref{fig:stochastic} shows spectra, spacial correlation functions and the PDFs at the statistically steady states for Eq. (\ref{Eq021}) for different coefficients $p_{2}$ of the stochastic pumping term (\ref{Eq051}) and with all other parameters (\ref{parameters}) fixed. The same functions for pure deterministic pumping system $p_{2}=0$ coincide with that for small stochastic pumping $p_{2}=10^{-4}$ and are not shown on FIG.~\ref{fig:stochastic}.

An evident result of superimposed stochastic pumping is a mild widening of spectra, nevertheless normalized spacial correlation functions turn out to be almost identical with each other. Changes in the PDFs first become noticeable at sufficiently large values of coefficient $p_{2}=30$ corresponding to average stochastic pumping amplitude $\sqrt{\langle|\Phi_{2}|^{2}\rangle_{\xi}}\approx 3$. Note that the PDFs on FIG.~\ref{fig:stochastic}c are presented versus non-renormalized squared amplitude $|\Psi|^{2}$. Therefore, in the statistically steady state statistics of waves for Eq. (\ref{Eq021}) is exceptionally stable against stochastic forcing. 

Dynamics of the system before the arrival to the statistically steady state is also perturbed by the addition of stochastic pumping: modulation instability develops much faster, oscillations of kinetic, four-wave and higher-order interactions energy cease earlier as well as the fluctuations and the existance of the peak at zeroth harmonic in spectra and the corresponding nonzeroth level for spacial correlation functions at infinity. Overall, spectra, spacial correlation functions and the PDFs significantly faster approach to their final shapes at the statistically steady states.

For very large stochastic pumping the dynamics of the system will be determined not by its interior structure, but by the stochastic pumping itself that in the sense of Eq. (\ref{Eq051}) should lead to Rayleigh PDFs. Indeed, for $p_{2}=100$ and average amplitude of the stochastic forcing term $\sqrt{\langle|\Phi_{2}|^{2}\rangle_{\xi}}\approx 10$, the PDF for Eq. (\ref{Eq021}) turns out to be very close to Rayleigh one. Statistics for for Eq. (\ref{Eq021}) with pure stochastic forcing turns out to be very similar to the latter case $p_{2}=100$ and is not presented here.

\begin{figure}[h] \centering
\includegraphics[width=130pt]{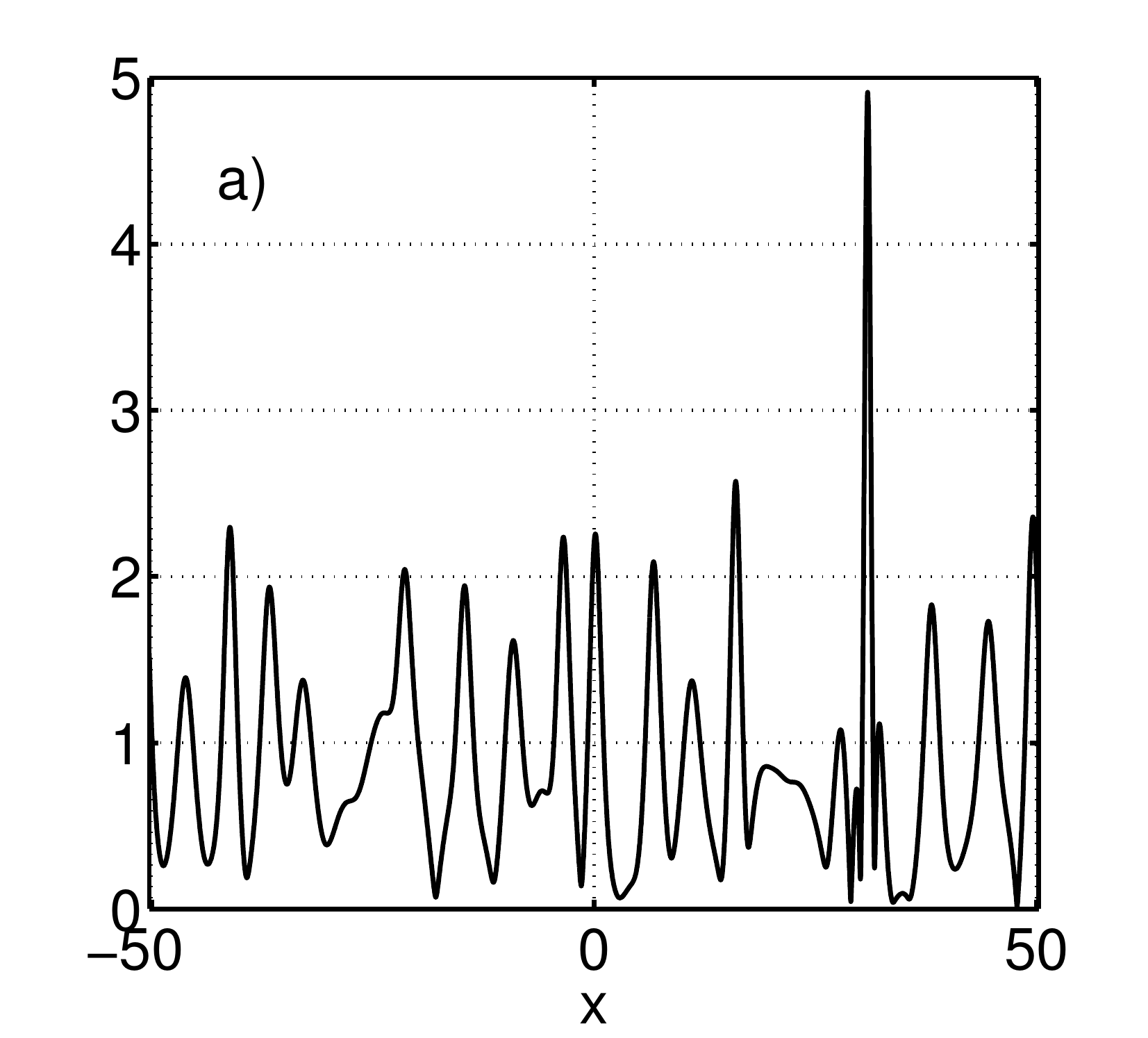}
\includegraphics[width=130pt]{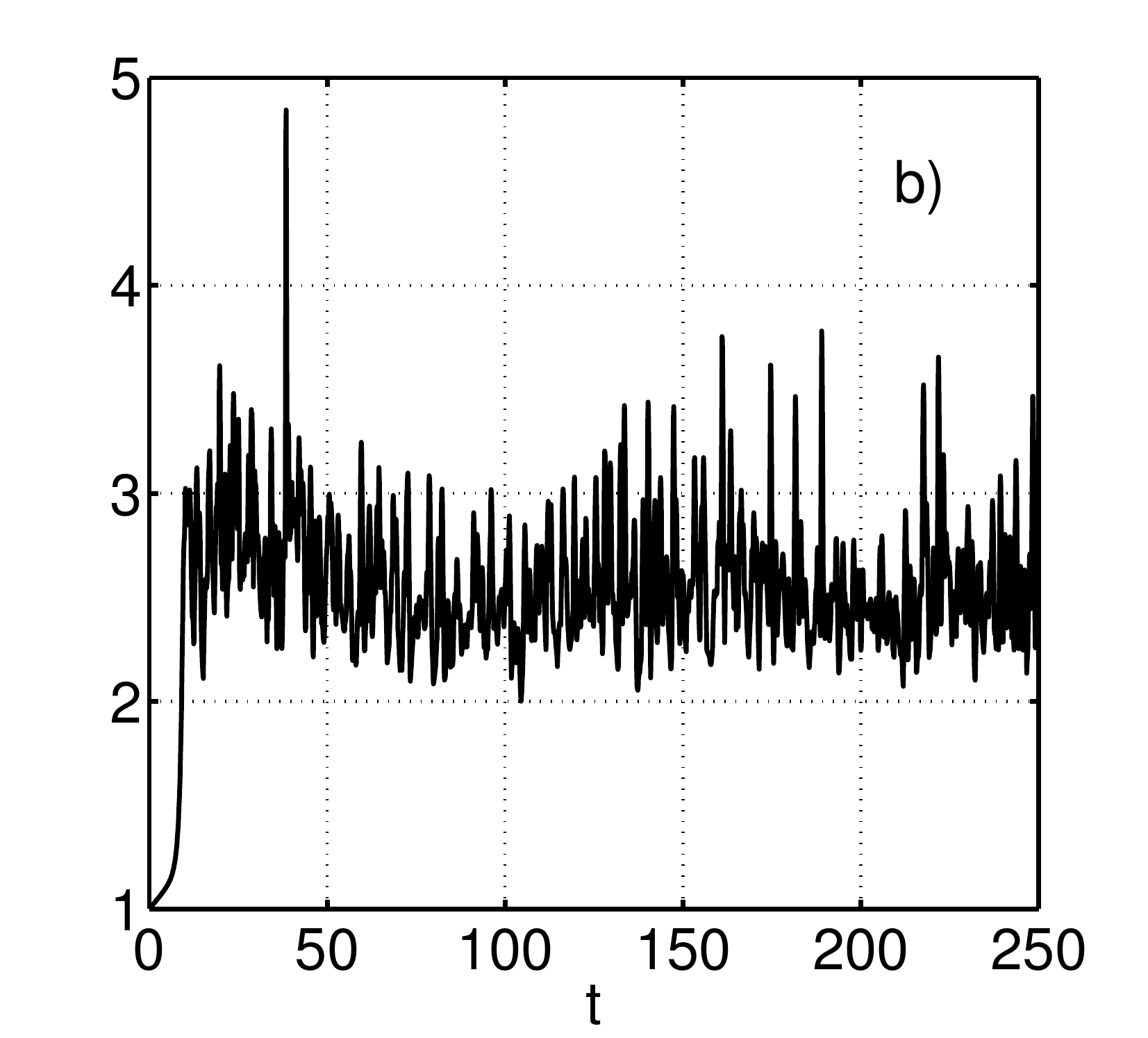}
\includegraphics[width=130pt]{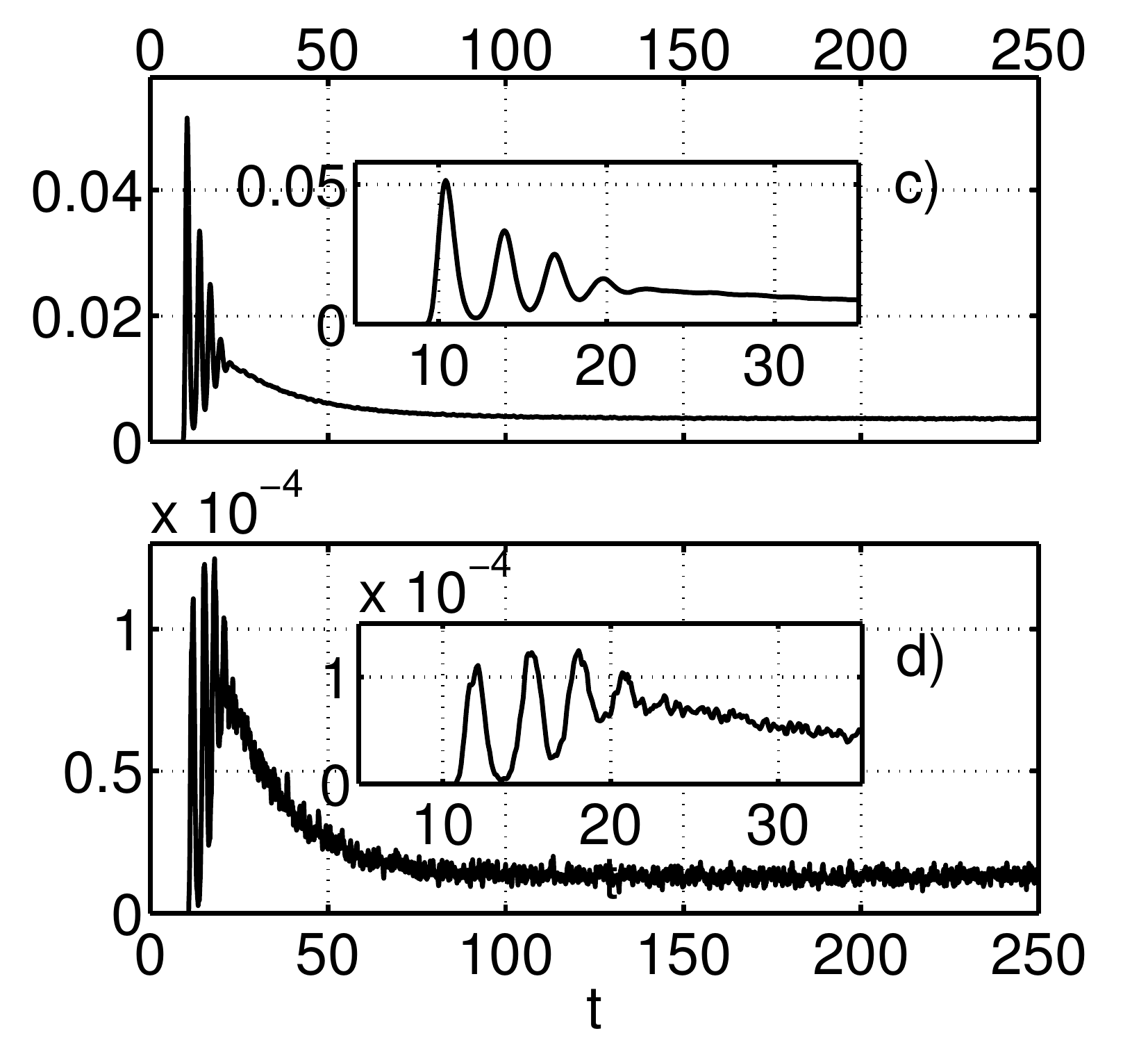}

\caption{\small {\it Field distribution $|\Psi|$ of a typical large wave event (a), evolution of $\max|\Psi|$ for the same run (b), and evolution of the frequencies of squared amplitudes appearance calculated as averaged over ensemble relative number of points where squared amplitude $|\Psi|^{2}$ exceeded thresholds $A_{1}^{2}=6$ (c) and $A_{2}^{2}=12$ (d) for Eq. (\ref{Eq021}) with parameters (\ref{parameters}). Small difference between the maxima of graphs (a) and (b) is attributed to the measurements procedure: evolution of $\max|\Psi|$ was gathered with significantly less temporal resolution than it was done for the search of a large wave event itself. Insets on graphs (c) and (d) demonstrate initial parts of graphs (c) and (d) in higher resolution.}}
\label{fig:large_wave_event}
\end{figure}

A typical large wave event for Eq. (\ref{Eq021}) with parameters (\ref{parameters}) as well as the evolution of absolute maximum $\max|\Psi|$ for the same run are shown on FIG.~\ref{fig:large_wave_event}a,b. Extreme events for Eq. (\ref{Eq021}) are large waves that appear for a very short period of time ($\Delta T\approx 1$ for event shown on FIG.~\ref{fig:large_wave_event}a,b) and then disappear. This is very different from optical rogue waves \cite{Dudley1, Dudley2, Taki1, Dudley3, Taki2}, that are large quasi-solitons travelling without significant changes in shape.

It is also possible to estimate frequencies of squared amplitudes appearance by calculation of averaged over ensemble relative number of points where squared amplitudes $|\Psi|^{2}$ exceed given thresholds $A^{2}$; such frequencies for thresholds $A_{1}^{2}=6$ and $A_{2}^{2}=12$ are shown on FIG.~\ref{fig:large_wave_event}c,d. In the nonlinear stage of modulation instability these frequencies oscillate with time with almost opposite phases very similar to the integrable case (\ref{Eq01}) studied in \cite{Agafontsev2}, but the oscillations cease significantly faster to $t\sim 25$ and then the frequencies monotonically decrease to some constant levels achieved in the statistically steady state. The frequencies are in average higher for the period of time from the nonlinear stage of modulation instability to the arrival into the statistically steady state $t\in[10, 100]$ becase average squared amplitude $\langle|\Psi|^{2}\rangle=\langle N\rangle/\int dx$ and wave action $N$ for this time interval are higher (
compare to FIG.~\ref{fig:EnergyEvolution}b, see also FIG.~\ref{fig:SCPevolution}d where the effect of different average squared amplitudes is eliminated).

\begin{figure}[h] \centering
\includegraphics[width=130pt]{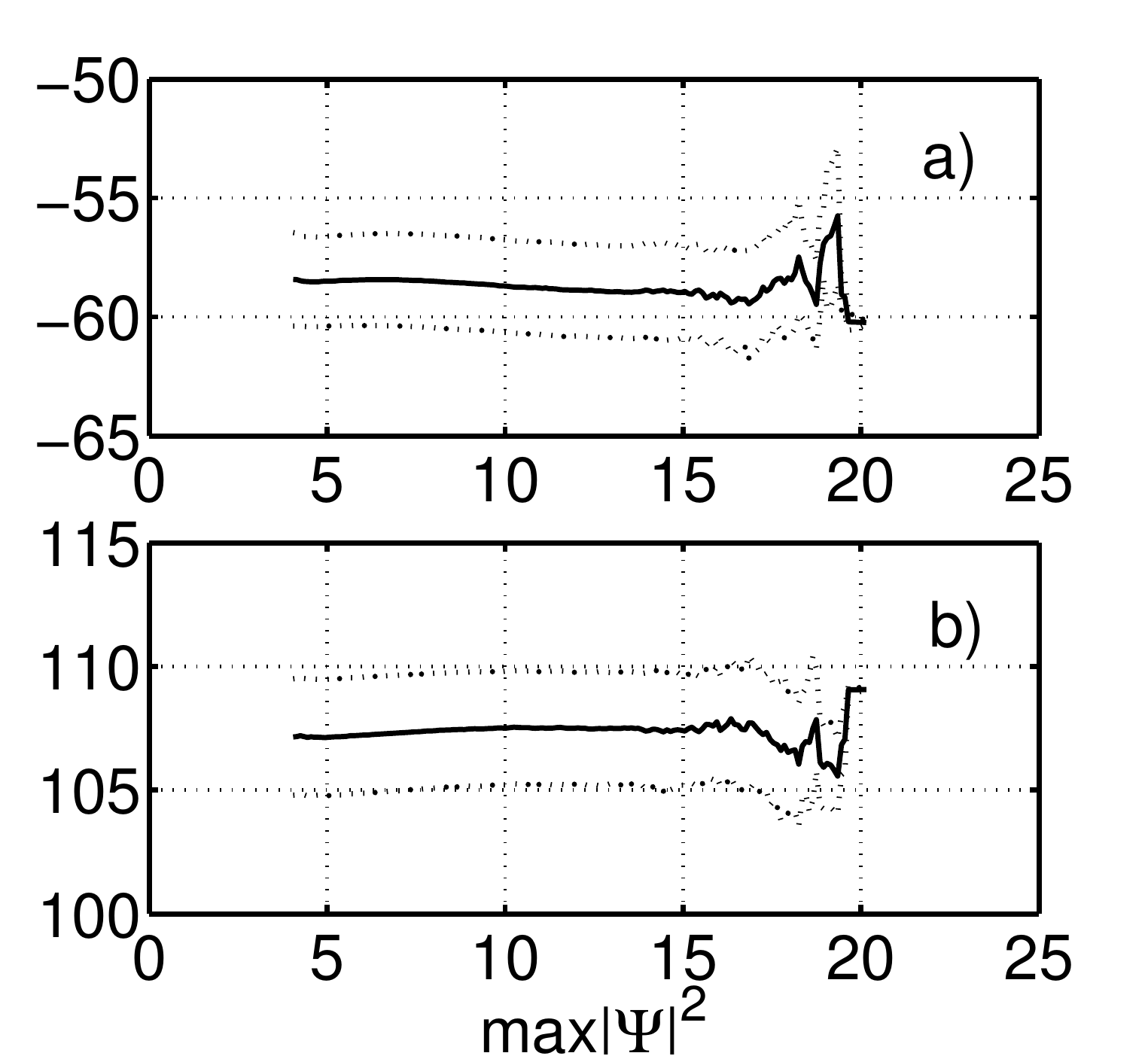}
\includegraphics[width=130pt]{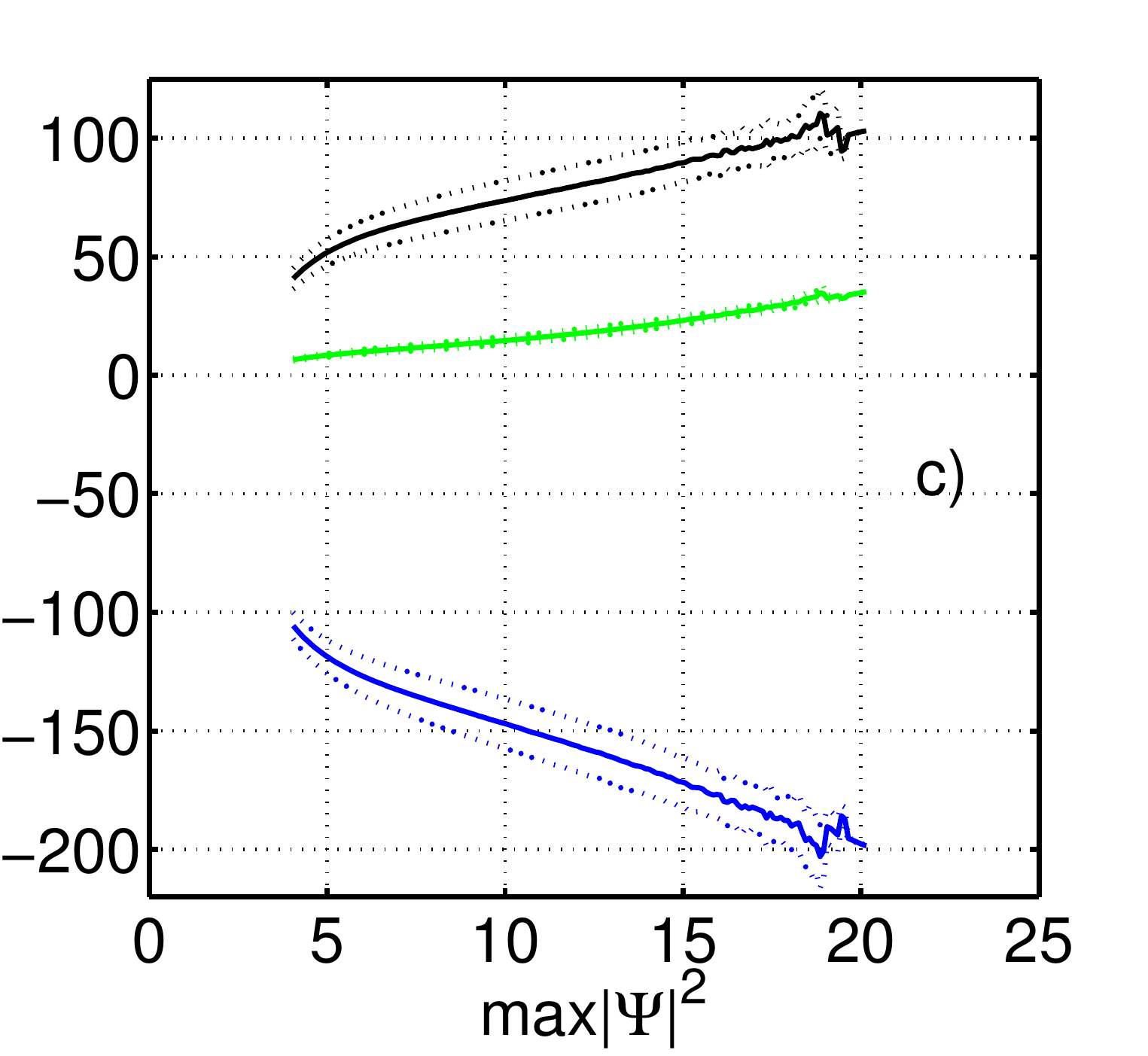}

\caption{\small {\it  (Color on-line) Averaged over distributions with the given squared absolute maximum $\max|\Psi|^{2}$ integral characteristics of wave field $\Psi$, also averaged over ensemble and time $t\in[200, 250]$, depending on $\max|\Psi|^{2}$: (a) total energy $\langle E\rangle_{\max|\Psi|^{2}}$, (b) wave action $\langle N\rangle_{\max|\Psi|^{2}}$ and (c) kinetic energy $\langle H_{d}\rangle_{\max|\Psi|^{2}}$ (black), four-wave interactions $\langle H_{4}\rangle_{\max|\Psi|^{2}}$ (blue) and higher-order interactions energy $\langle H_{6}\rangle_{\max|\Psi|^{2}}$ (green) for Eq. (\ref{Eq021}) with parameters (\ref{parameters}). Solid lines - mean values, dashed lines - borders for the corresponding standard deviations.}}
\label{fig:EnergyHistograms}
\end{figure}

Since Eq. (\ref{Eq021}) is a dissipative system, a situation is possible when the different parts of the PDFs corresponding to small, medium and high waves are composed of the distributions with significantly different integral characteristics, such as total energy and wave action. For example, distributions that compose far tails of the PDFs may have in average significantly higher wave action and thus average squared amplitude. In order to study this possibility we measure integral characteristics related to wave field $\Psi$, averaged over such distributions $\Psi$ that have the given squared absolute maximum $\max|\Psi|^{2}$; averaging in this way is signed as $\langle ..\rangle_{\max|\Psi|^{2}}$. As shown on FIG.~\ref{fig:EnergyHistograms}a,b, it turns our that both wave action and total energy virtually do not depend on the absolute maximum, i.e. all parts of the PDFs are composed of the distributions with the same mean squared amplitude, wave action and total energy. Concerning total energy, this 
result is significantly different from that for focusing six-wave interactions \cite{Agafontsev2} where total energy pronouncedly increased with $\max|\Psi|^{2}$.

Kinetic energy, four-wave interactions and higher-order interactions energy for Eq. (\ref{Eq021}) significantly increase in absolute values with $\max|\Psi|^{2}$ (FIG.~\ref{fig:EnergyHistograms}c). The latter circumstance is straightforward: higher absolute maximums significantly increase $|H_{4}|$ and $|H_{6}|$ directly, and also $|H_{d}|$ through higher gradients. It is noteworthy that absolute values of kinetic and four-waves interactions energy increase with $\max|\Psi|^{2}$ almost linearly.\\


{\bf 4.} We would like to underline the following of our results. First, we repeat one of our conclusions previously published in \cite{Agafontsev2}: presence of nonlinearity and significantly nonlinear regime of a system do not necessarily mean non-Rayleigh PDFs as demonstrated by the PDF for Eq. (\ref{Eq021}) in the absence of saturated nonlinearity $\alpha=0$ (inset on FIG.~\ref{fig:Palpha}a).

Second, evolution of spectra, spacial correlation functions and the PDFs for Eq. (\ref{Eq021}) is very similar to that for nonintegrable NLS equation accounting for small dumping and pumping terms studied in \cite{Agafontsev2}: kinetic, four- and higher-order interactions energy, as well as the frequencies of squared amplitudes appearance, oscillate with time in the nonlinear stage of modulation instability. At the same time energy spectrum has peak at zeroth harmonic $k=0$ and the corresponding spacial correlation function decays to some non-zero level; both the peak and the level and also the PDFs significantly fluctuate with time. Fluctuations and oscillations gradually cease to some universal time shift and simultaneously the peak in spectra and the nonzeroth level for the correlation functions disappear; then the system approaches to its statistically steady state when high waves appear randomly in space and time. 

Starting from some critical levels of saturation parameter $\alpha$ and average amplitude $\sqrt{\langle|\Psi|^{2}\rangle}$ achieved with some dumping and pumping coefficients, for the statistically steady states we observe power-law region on the PDFs for small and medium amplitudes - waves that occur in the system most frequently; power-law region is followed by intermediate region for higher amplitudes and then Rayleigh far tail. If power-law region is present, it turns out to be universal with respect to saturation, dumping and pumping parameters, $PDF(|\Psi|)\sim |\Psi|^{-1}$; it is more pronounced for higher saturation parameters and higher average amplitudes. Saturated nonlinearity is important for systems where extremely large waves are achieved, therefore waves from power-law region may still be very large in physical variables.

In the statistically steady states all parts of the PDFs, corresponding to small, medium and large waves, are composed of the distributions that have the same wave action, mean squared amplitude and total energy. Large wave events appear for very short time and then disappear. Correlation of phases becomes significant for high waves and increases their probability of occurrence approximately by one order of magnitude. 

Finally, statistics of waves for system (\ref{Eq021}) turns out to be exceptionally stable against additional stochastic forces as was checked by our experiments with superimposed deterministic and stochastic pumping terms. The PDFs start changing when stochastic pumping approaches to the same order of magnitude as conservative terms of Eq. (\ref{Eq021}) have. This promises a very good opportunity to observe our results in real physical systems.

D. Agafontsev thanks E. Kuznetsov and V. Zakharov for valuable discussions concerning this publication, M. Fedoruk for access to and V. Kalyuzhny for assistance with Novosibirsk Supercomputer Center. This work was done in the framework of Russian Federation Government Grant (contract No. 11.G34.31.0035 with Ministry of Education and Science of RF), and also supported by the program of Presidium of RAS "Fundamental problems of nonlinear dynamics in mathematical and physical sciences", program of support for leading scientific schools of Russian Federation, RFBR grant 12-01-00943-a and also Sergei Badulin RFBR grant 11-05-01114-a.

\end{document}